# Star formation in interacting galaxy systems: UVIT imaging of NGC 7252 and NGC 5291

Geethika Santhosh,[1] Rakhi R,[1] Koshy George,[2] Smitha Subramanian,[3] and Indulekha Kavila[4]

[1]Department of Physics, N.S.S. College, Pandalam (Affiliated to University of Kerala), Kerala 689501, India
[2]Faculty of Physics, Ludwig-Maximilians-Universität, Scheinerstr. 1, 81679
[3] Indian Institute of Astrophysics, Bangalore 560034, India
[4]School of Pure and Applied Physics, Mahatma Gandhi University, Kottayam, Kerala 686560 India
Author for correspondence: , Email: rakhi@nsscollegepandalam.ac.in.

**Abstract**
Interactions play a significant role in the formation and evolution of galaxies in the Universe. The galaxy systems, NGC 7252 and NGC 5291 are two nearby interacting systems that are hosting Tidal Dwarf Galaxies (TDGs) and star-forming knots. The present work aims (a) To determine the attenuation-corrected star formation rate (SFR) of the interacting system NGC 7252 (b) To compare the star formation in the NGC 7252 system with that of the NGC 5291 system (c) To explore the relation between surface densities of gas and SFR in these two systems. The study utilises high-resolution FUV and NUV imaging data from the Ultraviolet Imaging Telescope (UVIT) on board AstroSat. Six star-forming regions, including the merger remnant, were identified in the NGC 7252 system. The SFR corrected for attenuation of the knots in the NGC 7252 system is determined using the continuum slope β calculated from the FUV-NUV colour. It has been observed that the attenuation-corrected SFR values of the knots in this system fall within the range of SFR values determined for the NGC 5291 knots. The TDGs in both systems adhere to the same Kennicutt-Schmidt (KS) relation as regular spiral galaxies.

**Keywords:** Galaxies: interactions – Galaxies: star formation – Ultraviolet: galaxies – Galaxies: dwarf – Galaxies: formation

## 1. Introduction

Interactions and mergers play a major role in the evolution of galaxies (Barnes and L. E. Hernquist 1991; Barnes and L. Hernquist 1996; Hopkins et al. 2008; Blumenthal and Barnes 2018; Kevin Schawinski et al. 2010; Yoon et al. 2022). Interactions between galaxies can explain the 'bridges' and 'tails' seen in disturbed galaxy pairs. They also lead to galactic mergers that often trigger bursts of star formation (Zwicky 1956; Toomre and Toomre 1972; Larson 1990; Barnes 1998).

Violent encounters between galaxies can change the morphology, Star Formation Rate (SFR), metallically and gas content, converting star-forming spiral galaxies to non-star-forming ellipticals (Schweizer and Seitzer 1992; Sugata Kaviraj et al. 2011). Interactions and mergers can trigger strong central starbursts and AGN activity in galaxies; during galaxy encounters, gravitational torques can remove the angular momentum from the gas in the disc systems resulting in large gas inflows into the galactic centers (Barnes and L. E. Hernquist 1991; Mihos and Hernquist 1996; Bournaud 2010). Compression of gas and dust clouds can speed up the rate of star formation. During interactions, tidal forces can generate long tidal structures like tails and bridges of gas, dust and stars. These tidal features, around the galaxies, may contain self-gravitating star-forming clumps of dwarf galaxy masses and sizes, known as Tidal Dwarf Galaxies (TDGs). TDGs are formed in situ from the gas and stellar matter that is pulled out of the discs of parent galaxies into the intergalactic space during interactions (Duc et al. 2000; Duc et al. 2007; Hancock et al. 2007; Hancock et al. 2009). Though TDGs resemble normal independent dwarf irregulars (dIrrs) and Blue Compact Dwarf (BCD) galaxies in the Universe (Duc and Mirabel 1999) in terms of their size, mass, and SFR, they are still different from normal dwarf galaxies. Compared to normal dwarf galaxies that are of primordial origin, TDGs are generally metal-rich and are considered to be free of dark matter (Duc and Mirabel 1999; Duc et al. 2000; Barnes and L. Hernquist 1992; Hunter, Hunsberger, and Roye 2000).

Numerical simulations (Barnes and L. Hernquist 1992; Elmegreen, Kaufman, and Thomasson 1993) as well as observations (Mirabel, Lutz, and Maza 1991; Mirabel, Dottori, and Lutz 1992; Duc and Mirabel 1994; Yoshida, Taniguchi, and Murayama 1994; Duc and Mirabel 1998; Duc et al. 2000; Boquien et al. 2009; Schechtman-Rook and Hess 2012; Sengupta et al. 2014; Sengupta et al. 2017) point to the possibility of dwarf galaxy formation during galaxy interactions and collisions. Thus, there are two possible scenarios for the formation of dwarf galaxies in the Universe: the formation from collapse of primordial gas clouds and tidal dwarf formation from galaxy-galaxy interactions (Hunter, Hunsberger, and Roye 2000). According to Hunsberger, Charlton, and Zaritsky (1996), galaxy interactions produce at least one-third to one-half of dwarf galaxies in compact groups. Sugata Kaviraj et al. (2012) estimated that ∼6 percent of dwarf galaxies in cluster environments could have a tidal origin.

### 1.1 The post merger galaxy NGC 7252

The NGC 7252 galaxy (also known as Arp 226 or Atoms-for-Peace galaxy) is a post-merger system formed as a result of the merger of two massive gas-rich disc galaxies having similar masses. Each galaxy has an approximate disc mass of $1.87 \times 10^{10}$ M$_\odot$ (Chien and Barnes 2010). Observations reveal that NGC 7252 possesses a single bright nucleus; both the light distribution and the inner gas disc are centered on this nucleus (Schweizer 1982). TDGs are observed at the tip of both the two tidal tails of NGC 7252, and a number of small star-forming knots are seen at the base of the tails, close to the remnant. A preliminary study on star formation in the NGC 7252 system is given in George, Joseph, Côté, et al. (2018) where they estimated the SFR (uncorrected for internal attenuation) of the knots and TDGs in the system.

### 1.2   The NGC 5291 interacting system

The NGC 5291 system is another system, that is considered to have formed due to a violent galaxy-galaxy collision. The NGC 5291 interacting system lies at the edge of the Abell 3574 cluster, and it consists of an early type galaxy (ETG) called NGC 5291 (morphological type: E/S0) and a companion galaxy known as Seashell. A huge collisional HI ring surrounds the two interacting galaxies.

The NGC 5291 ring structure hosts several young star-forming knots, several TDG candidates and 3 bonafide TDGs. These TDG candidates are formed in the HI ring due to collision with a high-velocity impactor (Bournaud et al. 2007); being of different origin, they may not strictly be called TDGs. However, for simplicity, the term TDG is used to refer to these dwarfs. A detailed study of the star-forming knots and TDGs in the NGC 5291 interacting system, made by making use of FUV and NUV data of the NGC 5291 system from UVIT, is presented in Rakhi et al. (2023).

Several interacting systems have been identified in the local Universe that are observed to have dwarf-sized objects identified as candidate TDGs in collisional debris/tidally drawn out material around them. However, it is possible that these objects may either be pre-existing dwarf entities or objects just apparently associated with the interacting system due to projection effects. The observed TDGs in the NGC 5291 and NGC 7252 systems are genuine condensations of gas and stars, have a local potential well, and possess higher metallicities (near solar metallicities) than that observed for typical independent dwarf galaxies, confirming that they are bonafide TDGs formed during the interaction process (Bournaud et al. 2007; Lelli et al. 2015).

Investigation of ongoing star formation due to merger events and other galaxy interactions in the nearby Universe is of great importance as it gives insights into the formation of galaxies as well as their evolution. Galaxy-galaxy collisions and mergers involving disc galaxies can result in the transformation of disc systems to early-type galaxies (S0/E). Both NGC 7252 and NGC 5291 central galaxies are observed to be ETGs (Longmore et al. 1979; Schweizer 1982; Hibbard et al. 1994). Interactions between galaxies can drive AGN activity. An AGN has been discovered by optical spectroscopy in the galaxy NGC 5291 (Ingyin Zaw, Y.-P. Chen, and Glennys R. Farrar 2019; I. Zaw, Y. .-. Chen, and G. R. Farrar 2019). Studies have identified signatures of low luminosity AGN activity in the center of the NGC 7252 remnant (Weaver et al. 2018; George, Joseph, Mondal, et al. 2018). Li et al. 2023 describes the AGN activity to be fading in the post merger remnant phase.

### 1.3   Ultraviolet Observations

The ultraviolet (UV) continuum is used as a direct tracer of recent star formation. Massive stars (O, B, A types) which emit UV radiation copiously have short lifetimes, typically below $\sim$ hundred million years (Kennicutt and Evans 2012; Calzetti 2013). FUV flux and NUV flux, both are sensitive to stars in the age range of about 0-100 Myr and 0-200 Myr, respectively (Kennicutt and Evans 2012, Table 1). Thus, we can obtain information about the recent star formation activity of extra-galactic systems by examining their deep UV images (Kennicutt and Evans 2012). Since galaxy-galaxy collisions trigger star formation activity in the central regions of interacting galaxies/merger remnants and along their tidal features, interacting systems are good laboratories to observe and study such star formation activity which is considered to be more frequent in the earlier phases of the Universe. With the launch of the Ultraviolet Imaging Telescope (UVIT) on board AstroSat, a stream of new high-resolution observational information on star formation has emerged in the past few years (George, Joseph, Côté, et al. 2018; Mondal, Subramaniam, and George 2018, 2019; Poggianti et al. 2019; Mondal, Subramaniam, and George 2021; Mondal et al. 2021; Hota et al. 2021; Ujjwal et al. 2022; Joseph et al. 2022; Mahajan et al. 2022; George 2023; Robin et al. 2024).

In the present paper, we compare star formation activity in the interacting galaxy systems, NGC 5291 and NGC 7252. The sample selection is influenced by the fact that the two systems have intense star formation in their interaction debris and have bonafide TDGs. High-resolution UVIT NUV and FUV data are available, and both systems are observed through the same NUV and FUV filters. The distances to the two systems are roughly the same. Also, though their formation scenarios are different, substructure formation in the tidal tails of NGC 7252 and the collisional ring of NGC 5291 is observed to be similar, justifying a comparison of the star formation in their interaction debris. Attenuation-corrected SFR is determined with the help of observations made with UVIT.

The paper is structured as follows: Section 2 describes the details of data reduction, source extraction and identification of star-forming knots. Results are presented in Section 3 while Section 4 presents the discussion on the results. Conclusions are presented in Section 5.

Throughout the paper, $\Lambda_{CDM}$ cosmology with Hubble parameter, $H_0$ = 71 km s$^{-1}$ Mpc$^{-1}$, $\Omega_M$ = 0.27 and $\Omega_\Lambda$ = 0.73 (Komatsu et al. 2011) is adopted.

## 2.   Data & Analysis

### 2.1   Data

The post-merger galaxy, NGC 7252 and the interacting system NGC 5291 were observed with the Ultraviolet Imaging Telescope (UVIT) on board AstroSat.

**Table 1.** NGC 7252 and NGC 5291: UVIT observations

| System | RA | Dec | Distance (Mpc) | Linear distance corresponding to 1″ (kpc) | Channel | Filter name | $\lambda_{mean}$ (Å) | $\Delta\lambda$ (Å) | Integration time (s) |
|---|---|---|---|---|---|---|---|---|---|
| NGC 7252 | $22^h20^m44.75^s$ | $-24^040'41.75''$ | 68 | 0.32 | FUV | F148W | 1481 | 500 | 8138 |
|  |  |  |  |  | NUV | N242W | 2418 | 785 | 7915 |
| NGC 5291 | $13^h47^m24.48^s$ | $-30^024'25.20''$ | 62 | 0.30 | FUV | F148W | 1481 | 500 | 8242 |
|  |  |  |  |  | NUV | N242W | 2418 | 785 | 8079 |

NOTE: $\lambda_{mean}$ and $\Delta\lambda$ respectively are the effective wavelength and bandwidth of the filters. Details on the UVIT filters can be found in (Tandon et al. 2017).

The details of the observations are summarised in Table 1. Figure 1 shows the NUV images of NGC 7252 and NGC 5291 systems.

### 2.2 Data Reduction

Both NGC 7252 and NGC 5291 Level 1 (L1) UVIT data are reduced to Level 2 (L2) science ready images using CCDLAB pipeline (Postma and Leahy 2017, 2021). UVIT data is flat-fielded and corrected for fixed pattern noise, distortion, and drift using CCDLAB. The orbit-wise images are aligned to a common frame before merging. CCDLAB provides an option to optimise the Point Spread Function (PSF) of the source to get the best PSF, which corresponds to an improved, narrower radial profile (Postma and Leahy 2021). The NUV and FUV master images are PSF optimised ones. To optimise the PSF, CCDLAB pipeline analyses time-bins of photon centroids about the mean positions of their corresponding sources. In effect, this generates a residual drift series which can then be used to optimise the PSF (Postma et al. 2023). The automated WCS solver in CCDLAB (Postma and Leahy 2020) is then used to align the images with respect to the sky coordinates. The final NUV and FUV master images have an array size of 4096 × 4096 pixels where the size of a single square pixel corresponds to $0.416''$ in the sky. We computed the FWHM for 9 stars near to the galaxy by fitting a Gaussian to the light profile. The median value of which is then taken as the FWHM of the PSF of the imaging data. The final PSF of the NUV and FUV images thus obtained are $\sim 1.10''$ and $\sim 1.34''$ respectively.

Since both the NGC 7252 remnant and the NGC 5291 galaxy have regions with emission from AGN at the center (Weaver et al. 2018; I. Zaw, Y. .-. Chen, and G. R. Farrar 2019), source extraction is performed after masking the AGN in the images. For the NGC 7252 remnant, George, Joseph, Mondal, et al. (2018) estimated the size of the AGN-dominated region to be 1.3 kpc. However, there is no such estimate available for the NGC 5291 galaxy. Hence, for both the galaxies, the AGN contribution to the flux is removed by masking the central regions using an aperture with a diameter of 1.3 kpc. All fluxes provided in the present work are thus the values obtained after removing the AGN contribution.

### 2.3 Source extraction and identification of the knots in NGC 7252

Source extraction is performed using the ProFound source extraction package (Robotham et al. 2018). The package includes a suite of low, mid, and high-level functions for simple and advanced source extraction[a]. The highest level ProFound function, *profoundProFound*, detects sources, generates a segmentation map, and extracts photometry. The function detects pixels from the input image that are above a given threshold limit and uses a watershed de-blending algorithm for deblending the pixels. The deblended pixels make up a segment, and each segment corresponds to a source in the image. The segments are dilated iteratively until flux convergence is reached. The function generates a dilated segmentation map that shows the extent of each segment. The photometry is extracted from this map. Since the function does not assume a fixed aperture for extracting photometry, rather performs photometric extraction from the dilated segments, the ProFound source extraction package is most suitable for the extraction of flux from the sources with complex morphology (Robotham et al. 2018).

The function *profoundProFound* performs well with the default parameters (See the documentation[b] for the details of input parameters).The values of the input parameters, viz., sky (mean sky background in image units), skyRMS (standard deviation of the sky pixels), skycut (the minimum threshold for detecting pixels in units of skyRMS), sigma (the smoothness parameter used for generating segmentation map), pixcut (the minimum number of pixels for object identification), magzero (magnitude zero point of the filter) are optimised for better image segmentation, to ensure a signal to noise ratio (SNR) > 5 for the detected sources and to get accurate source statistics including flux and magnitudes. Sources are extracted from the coarser resolution FUV image using the *profoundProFound* function and the FUV image segmentation map generated by the function is used for extracting photometry from the NUV image by forcing the FUV segmentation map on the NUV image.

The star-forming knots in the NGC 7252 system are identified by overlaying the ProFound-extracted sources on the FUV image and selecting those sources corresponding to the tidal tails and the main body of the NGC 7252 system through

---

a. https://www.rdocumentation.org/packages/ProFound/versions/1.14.1
b. https://www.rdocumentation.org/packages/ProFound/versions/1.14.1/topics/ProFound

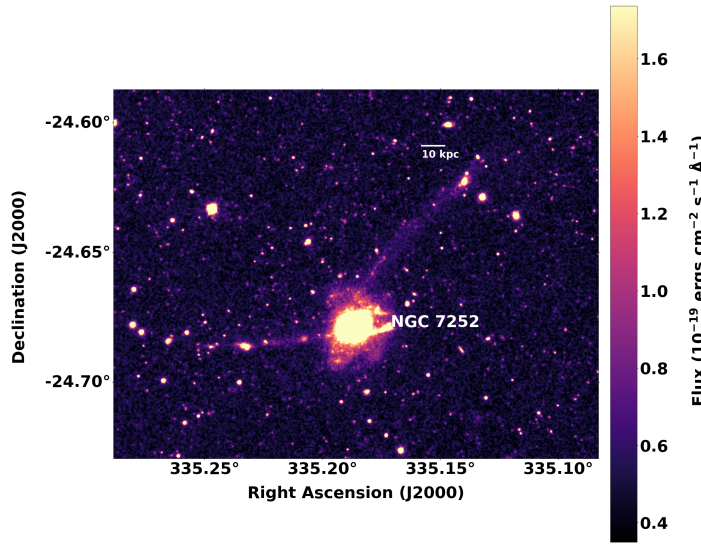

(a)

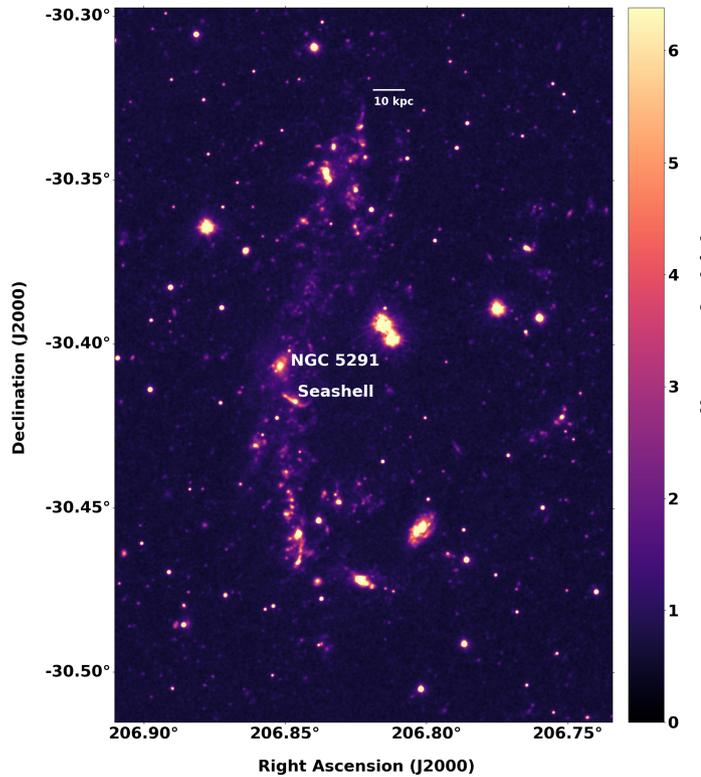

(b)

**Figure 1.** UVIT NUV images of the (a) NGC 7252 and (b) NGC 5291 systems. North is up and east is towards the left of the image. The colour scale is in units of $erg/s/cm^2/Å$. (a) The NGC 7252 system: Two tidal tails, the eastern tail and the north-western tail, extend from the NGC 7252 remnant (labeled NGC 7252). Star-forming knots are visible along these tails. (b) The NGC 5291 system: Several star-forming knots extend toward the north, south and west tracing a fragmented ring structure.

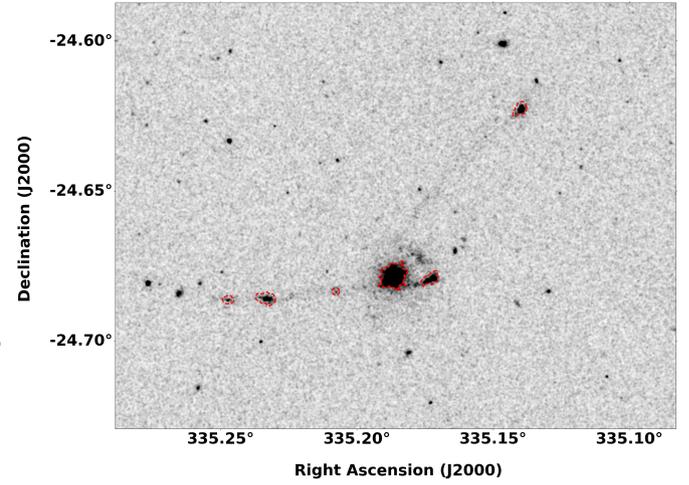

**Figure 2.** NGC 7252 FUV image with segment contours overlaid

a visual inspection. The selected sources are further compared with the Dark Energy Camera Legacy Survey (DECaLS) image of the NGC 7252 system. The DECaLS images the sky in g, r and z optical bands using the telescope at the Cerro Tololo Inter-American Observatory (Dey et al. 2019). The DECaLS image of the NGC 7252 system from the Data Release 10 (DR10) of the survey is utilised in the present study for the confirmation of knots. We performed a visual inspection of DECaLS colour composite images and confirmed that the six selected regions (including the merger remnant) appear bright blue. For NUV and FUV images, flux calibration is carried out using the zero point and unit conversion factors given in Tandon et al. (2017) and updated in Tandon et al. (2020).

George, Joseph, Côté, et al. (2018) studied six star-forming regions in the NGC 7252 post-merger system and estimated their star formation rates (uncorrected for internal attenuation). In the present study, we have identified these star-forming regions (marked as A, B, C, D, E, H in Figure 3). These regions are considered for further analysis.

### 2.4 Source extraction and identification of the knots in NGC 5291

The steps involved in the extraction of the sources and the identification of star-forming knots are explained in detail by Rakhi et al. (2023). A total of 206 sources were extracted from the FUV image by the ProFound source extraction program. The FUV segmentation map was then forced on the NUV image to extract the NUV photometric parameters. To identify the knots belonging to the NGC 5291 system, the FUV–NUV colour distribution of the extracted sources was examined, and those sources that fall within one standard deviation from the mean were considered. Those sources that fall within the HI contour were then compared with the DECaLS optical image and, a total of 57 knots which were bright blue were selected for detailed analysis.

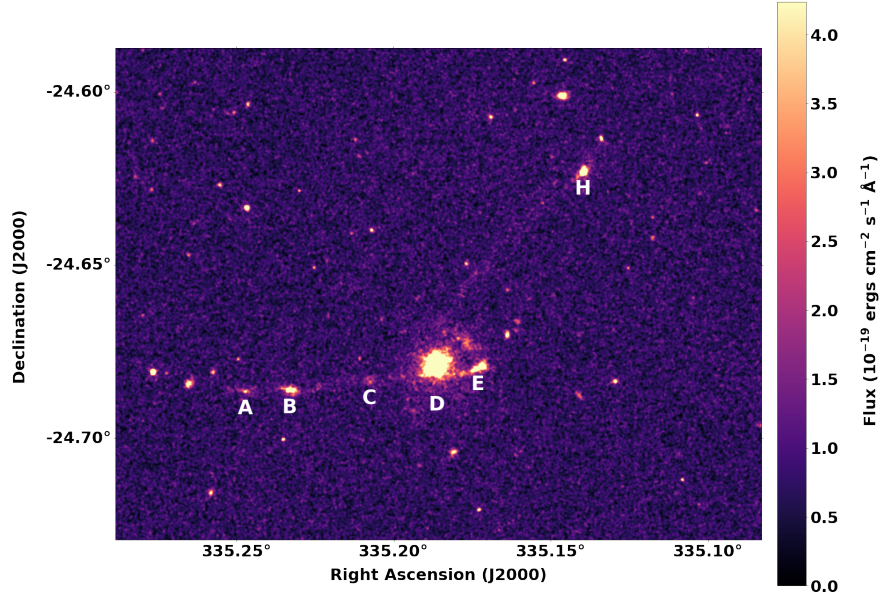

**Figure 3.** NGC 7252 FUV image with selected knots marked. D is the main body NGC 7252 remnant. North is up and east is towards the left of the image.

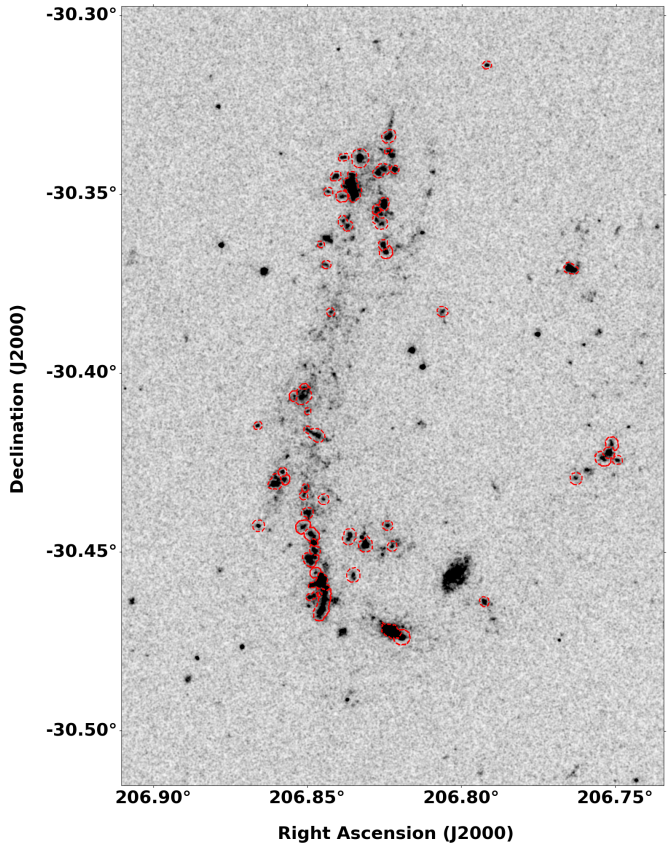

**Figure 4.** NGC 5291 FUV image with segment contours overlaid.

### 2.5 Correction for Galactic extinction

Galactic extinction is estimated from the Cardelli extinction law (Cardelli, Clayton, and Mathis 1989) by taking $R_v$ = 3.1.

For the UVIT FUV ($\lambda_{mean}$=1481 Å) and NUV ($\lambda_{mean}$=2418 Å) bands, Galactic extinction is given by the following equations respectively:

$$A_{FUV}\,(Galactic) = 8.34 \times E(B-V) \quad (1)$$

$$A_{NUV}\,(Galactic) = 7.75 \times E(B-V) \quad (2)$$

We use the E(B-V) value of 0.0259 ± 0.0002 mag from the Schlafly and Finkbeiner (2011) reddening map[c] for the computation of Galactic extinction in the direction of NGC 7252. The foreground Galactic extinction thus estimated are: $A_{FUV}(Galactic)$= 0.216 mag and $A_{NUV}(Galactic)$= 0.201 mag. The Galactic extinction-corrected FUV and NUV magnitudes are further corrected for internal attenuation, which is explained in Section 3.1.

For the resolved knots in the NGC 5291 system (Rakhi et al. 2023), Galactic extinction in the FUV and NUV bands are estimated using Eq. 1 and Eq. 2 respectively; E(B-V) = 0.0543 ± 0.0013 from Schlafly and Finkbeiner (2011) reddening map. Both NGC 7252 and NGC 5291 systems are observed using the same FUV and NUV UVIT filters (see Table 1 for details). The estimated values for Galactic extinction are: $A_{FUV}$ (Galactic) = 0.453 mag and $A_{NUV}$ (Galactic) = 0.421 mag.

## 3. Results

### 3.1 Computation of attenuation for NGC 7252 and NGC 5291

One possible measure of dust attenuation in star-forming galaxies is the slope of the UV continuum (Boquien et al. 2012; Overzier et al. 2011) along with the Meurer relation [hereafter M99] (Meurer, Heckman, and Calzetti 1999). The UV

---

c. https://irsa.ipac.caltech.edu/applications/DUST/

Table 2. Estimated parameters of the star-forming regions in the NGC 7252 interacting system

| ID[a] | Area ($kpc^2$) | $SFR_{FUV}(uncorr)$[b] ($M_\odot$/yr) Present study | $SFR_{FUV}(Galactic)$[c] ($M_\odot$/yr) Present study | $SFR_{FUV}$ ($M_\odot$/yr) George et al. (2018a) | β | $A_{FUV}(Internal)$ (mag) | $SFR_{FUV}(corr)$[d] ($M_\odot$/yr) |
|---|---|---|---|---|---|---|---|
| A | 11.7 | $0.0043 \pm 0.0003$ | $0.0053 \pm 0.0004$ | 0.003 | $-1.58 \pm 0.17$ | $1.29 \pm 0.34$ | $0.017 \pm 0.006$ |
| B (TDG NGC7252E) | 30.4 | $0.019 \pm 0.001$ | $0.023 \pm 0.001$ | 0.018 | $-1.39 \pm 0.08$ | $1.67 \pm 0.16$ | $0.11 \pm 0.02$ |
| C | 6.5 | $0.0025 \pm 0.0003$ | $0.0031 \pm 0.0003$ | 0.002 | $-1.06 \pm 0.22$ | $2.33 \pm 0.43$ | $0.026 \pm 0.011$ |
| D (Remnant) | 81.8 | $0.36 \pm 0.003$ | $0.44 \pm 0.004$ | 0.658 | $-0.34 \pm 0.03$ | - | - |
| E | 25.6 | $0.037 \pm 0.001$ | $0.045 \pm 0.001$ | 0.045 | $-1.40 \pm 0.06$ | $1.64 \pm 0.12$ | $0.20 \pm 0.02$ |
| H (TDG NGC7252NW) | 22.7 | $0.029 \pm 0.001$ | $0.036 \pm 0.001$ | 0.034 | $-1.66 \pm 0.07$ | $1.12 \pm 0.13$ | $0.10 \pm 0.01$ |

NOTE: [a] ID as given in George, Joseph, Côté, et al. 2018; [b] Uncorrected SFR (uncorrected for both Galactic extinction and internal attenuation); [c] SFR corrected for Galactic extinction; [d] Corrected SFR (Corrected for both Galactic extinction and internal attenuation)

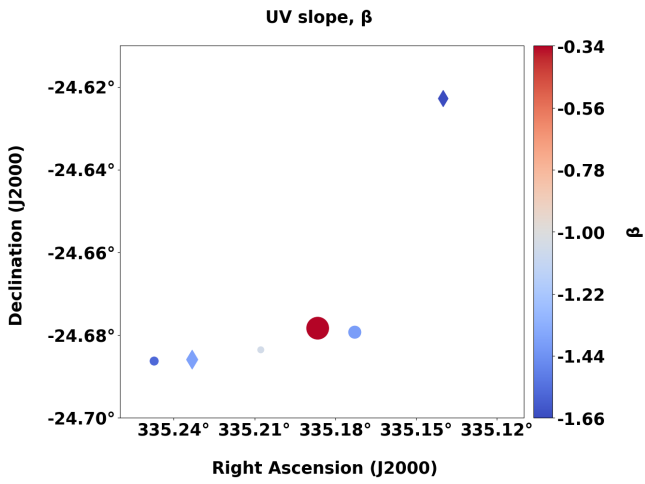

**Figure 5.** The β distribution of the star forming regions including the TDGs (diamonds) in the NGC 7252 system.

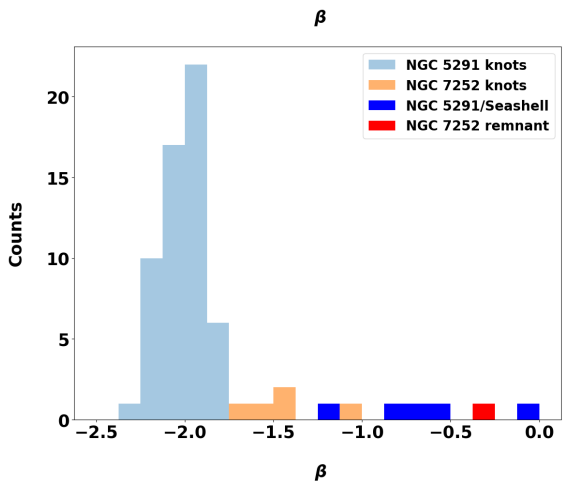

**Figure 6.** Distribution of β of the star forming regions in the NGC 7252 and NGC 5291 systems.

continuum of star-forming galaxies is characterised by the spectral index β with $f_\lambda \propto \lambda^\beta$ (Calzetti, Kinney, and Storchi-Bergmann 1994) for λ > 1200 Å; $f_\lambda$ (erg $cm^{-2}$ $s^{-1}$ $Å^{-1}$) is the flux density of the source. For UVIT FUV and NUV passbands, the slope of the UV continuum is given by:

$$\beta_{UVIT} = 1.88(m_{FUV} - m_{NUV}) - 2.0 \quad (3)$$

where $m_{FUV}$ and $m_{NUV}$ are the Galactic extinction-corrected FUV and NUV magnitudes respectively (Rakhi et al. 2023). The internal attenuation is calculated using the M99 relation (for starburst case):

$$A_{FUV} \text{ (Internal)} = 4.43 + 1.99 \beta \quad (4)$$

where β is given in Eq. 3.

The β values are estimated from the Galactic extinction-corrected FUV and NUV magnitudes of the star forming regions identified in the NGC 7252 and NGC 5291 systems and from the estimated β values, $A_{FUV}(Internal)$ values are estimated using Eq. 4. It is to be noted that, in the present study, the Galactic extinction-corrected magnitudes are considered for estimating β and $A_{FUV}(Internal)$ of star forming regions in the NGC 5291 system, whereas it was not considered in Rakhi et al. (2023). As a result, there is a difference of 0.12 mag, between the estimated values of $A_{FUV}(Internal)$ reported in this work and those presented in Rakhi et al. (2023).

We note that the choice of star formation history can affect the computed internal attenuation. As demonstrated in Boquien et al. (2012), using a starburst case can overestimate the attenuation and hence the computed SFR by an order of magnitude, compared to normal star formation. However, we used the same star formation history for all the knots in both galaxies studied in the present work, and hence, at the spatial scales probed by UVIT, the intrinsic β values of the knots could be slightly different from the values obtained here, due to slight differences between the knots in their star formation histories. This is particularly relevant for knots on the disks and at different distances from the center along the tidal tails.

The β values of the star-forming regions in the NGC 7252 interacting system are given in Figure 5. The β values of the knots range from -1.66 to -1.06 whereas the NGC 7252 post merger remnant has a β value of -0.34. For the knots in the NGC 5291 system, the values of β range from −2.26 to −1.79 while the β values of the resolved star forming regions of the galaxy main body of the NGC 5291 ranges from −1.17 to −0.08. For the Seashell galaxy, β ranges from −0.54 to 0.01. Figure 6 gives a comparison of the β values of the star forming regions (including the galaxy main bodies) of the NGC 7252 and NGC 5291 systems. The β values and hence the $A_{FUV}(Internal)$ derived from the β values are considerably higher for the main bodies of the two systems. The galaxy disk can contain stars of different stellar populations with the evolved population contributing to the UV continuum. To check this, we use NUV-r colour analysis, which is described in Section 3.2.

## 3.2 NUV-r colour analysis

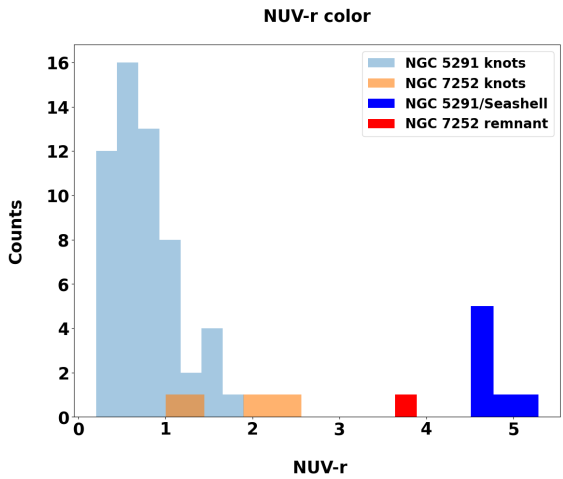

**Figure 7.** Distribution of NUV-r colour of the star-forming regions in the NGC 5291 and NGC 7252 systems. The regions with NUV-r > 3 are the resolved star forming regions of the galaxy main bodies.

The ultraviolet continuum can have a contribution from the evolved population of stars on the horizontal branch with ages > 8 Gyr. This is particularly true for the flux from the disk of galaxies where multiple populations of stars can be present. In collisional ring where in situ star formation is happening, there cannot be the presence of such a population. But if old stars are pulled from the disk there can be contribution from evolved population. We check for the presence of any evolved population using the NUV-r colour of the detected knots on the disk and collisional debris of both the systems. NUV-r < 5.4 is expected from a young population, while NUV−r > 5.4 is expected from an old population of stars (K. Schawinski et al. 2007; S. Kaviraj et al. 2007). The DECaLS DR10 r-band (effective wavelength = 6382.6 Å (Schlafly and Finkbeiner 2011)) images, of the NGC 5291 and NGC 7252 interacting systems, are used to estimate the r–band magnitudes of the identified star forming knots and the galaxy main bodies. The r-band images from the Dark Energy Camera (DECam) have a PSF FWHM ∼1.18" which is lower than the PSF FWHM of the UVIT FUV images. The photometry is extracted by forcing the FUV segmentation map extracted from ProFound on the r-band images. The r-band fluxes given in units of nanomaggies are converted to magnitudes using the conversion equation: $AB\,mag = 22.5 - 2.5 \times log_{10}(nanomaggy)$, given in the DECaLS fits image header. The magnitudes obtained are corrected for Galactic extinction following the O'Donnell extinction law (O'Donnell 1994) before estimating the NUV-r colour.

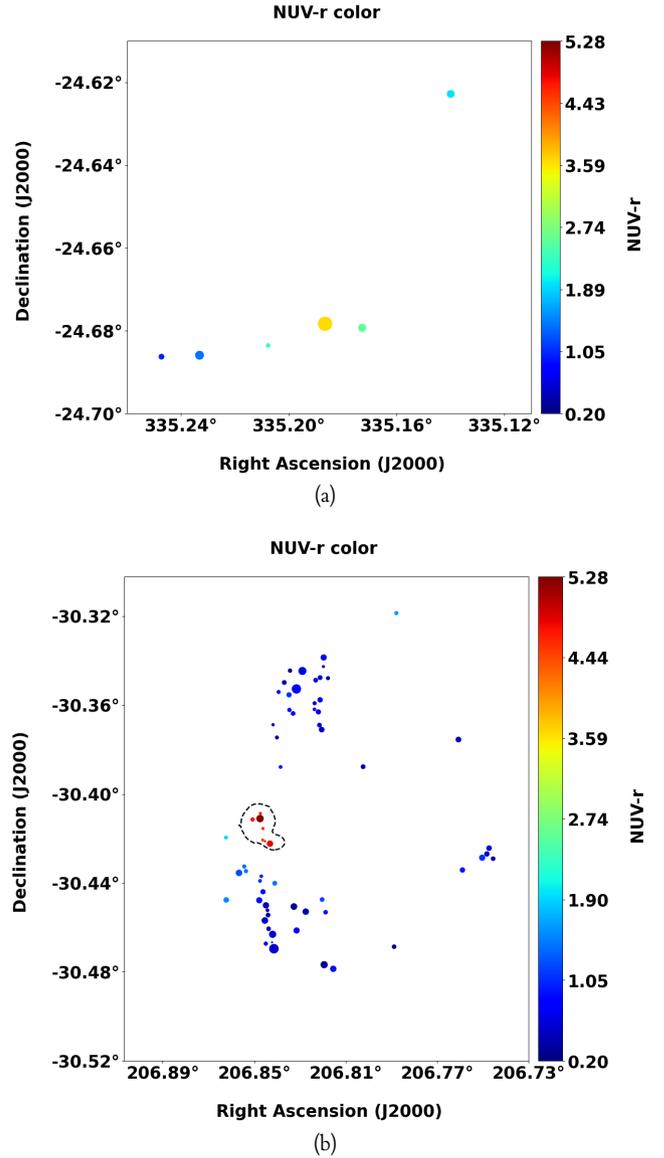

**Figure 8.** NUV-r colour of the star forming regions (including the main body) (a) NGC 7252 and (b) NGC 5291 systems. The markers are resized to show the relative sizes of the star-forming regions. Contour at level 22 mag/arcsec² from the z-band image is overlaid.

The histogram of NUV-r colour of the star-forming regions in the NGC 5291 and NGC 7252 systems are shown in Figure 7 and the NUV-r colour of the main body and the knots of the two systems are given in Figure 8. Contour corresponding to surface brightness level of 22 mag/arcsec$^2$ from the DECaLS z-band imaging data is displayed to trace the stellar disc of the NGC 5291 and Seashell galaxies (Figure 8b). It is observed that the regions coincident with the central interacting galaxies of the NGC 5291 system exhibit NUV-r colours between 4.5 and 5.28, while the identified knots in the NGC 5291 system show NUV-r colours between 0.20 and 1.9. For the NGC 7252 system, the NUV-r colour of the merger remnant is 3.64 while the NUV-r colours of the knots range from 1.00 to 2.56. All the star forming regions in both the systems exhibit NUV-r < 5.4 indicating star formation in the last 1-2 Gyr (recent star formation (RSF)) (K. Schawinski et al. 2007; S. Kaviraj et al. 2007). Though the NUV-r colours indicate no significant contribution to the UV continuum from hot evolved stars, there could be contribution to NUV emission by stars older than 100 Myr (Hao et al. 2011). This could be the reason for the higher values of β for the interacting galaxies of NGC 5291 and the merger remnant of NGC 7252, where the NUV-r colours are redder than 3.5 mag. We note that the β values and the extinction values of the main bodies of galaxies in NGC 5291 and the merger remnant of NGC 7252 could be over-estimated and hence internal attenuation is not estimated for main bodies in further analysis.

### 3.3 Star Formation Rate ($SFR_{FUV}$) of the NGC 7252 knots and the NGC 5291 knots

For the estimation of $SFR_{FUV}$, we use the following relation (Iglesias-Páramo et al. 2006; Cortese, Gavazzi, and Boselli 2008).

$$SFR_{FUV}[M_\odot/yr] = \frac{L_{FUV}[erg/sec]}{3.83 \times 10^{33}} \times 10^{-9.51} \quad (5)$$

where, $L_{FUV}$ is the luminosity. The flux density is converted to flux and to luminosity using the effective wavelength of the filter, $\lambda_{mean}$ and the distance to the source. This relation is based on the assumption of a constant rate of star formation over a timescale of $10^8$ years, with a Salpeter initial mass function (IMF) (Salpeter 1955) for stars with masses from 0.1 to 100 $M_\odot$ as described in Iglesias-Páramo et al. (2006) and in Cortese, Gavazzi, and Boselli (2008).

The attenuation is calculated from the slope of the UV continuum (β) derived from the FUV-NUV colours of each of the selected knots (see Eq. 3). The corrected SFR is then derived from the dust attenuation-corrected FUV fluxes.

It is to be noted that adopting a Milky Way-like IMF (Kroupa (2001) or Chabrier (2003) IMF), which has fewer low-mass stars compared to the Salpeter IMF, would tend to reduce our SFR estimates. Specifically, assuming a Kroupa IMF instead of the Salpeter IMF would decrease the SFR values by a factor of approximately 1.6 (Calzetti 2013).

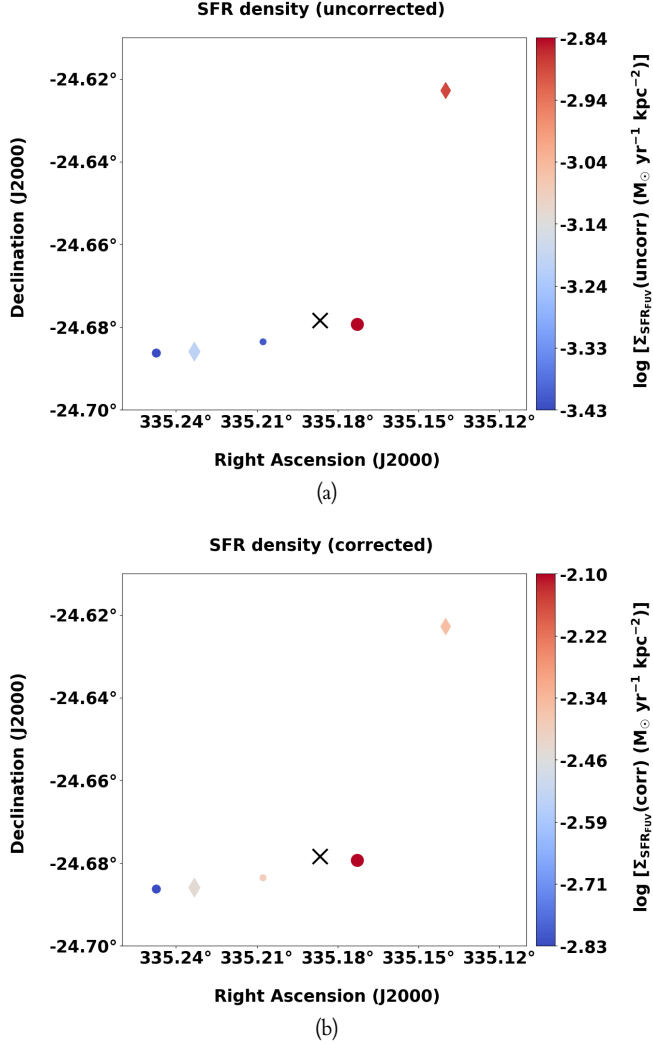

**Figure 9.** SFR surface density, $\Sigma_{SFR_{FUV}}$, across the location of the knots (circles) and TDGs (diamonds) in the NGC 7252 system. The black cross marks the location of the remnant. (a) $\Sigma_{SFR_{FUV}(uncorr)}$, uncorrected for Galactic extinction and internal attenuation and (b) $\Sigma_{SFR_{FUV}(corr)}$, corrected for both Galactic extinction and internal attenuation.

For the star-forming knots in the NGC 7252 system, the estimated values of β range from -1.66 to -1.06. The numerical value of the internal attenuation, $A_{FUV}(Internal)$ derived from β using the M99 relation for the knots ranges from 1.12 to 2.33. The integrated $SFR_{FUV}(corr)$ (SFR corrected for both Galactic extinction and internal attenuation) of the knots in the system is $0.46 \pm 0.03$ $M_\odot$/yr.

The estimate for the total $SFR_{FUV}(uncorr)$ (SFR without correction for both Galactic extinction and internal attenuation) for the knots, excluding the remnant, is $0.092 \pm 0.002$ $M_\odot$/yr while the integrated $SFR_{FUV}(Galactic)$ (SFR corrected only for Galactic extinction) for the knots, excluding the remnant, is $0.11 \pm 0.002$ $M_\odot$/yr. The SFR surface density, $\Sigma_{SFR}$ [d] across the location of the knots and TDGs in the NGC 7252 system is shown in Figure 9.

For the knots in the NGC 5291 system, the β values range from -2.26 to -1.79 and the $A_{FUV}(Internal)$ values range from 0.0 to 0.87 mag. The total $SFR_{FUV}(corr)$ of the knots in the NGC 5291 system is $2.4 \pm 0.06$ $M_\odot$/yr. The total $SFR_{FUV}(uncorr)$ of the knots is $1.0 \pm 0.005$ $M_\odot$/yr while the integrated $SFR_{FUV}(Galactic)$ is $1.6 \pm 0.008$ $M_\odot$/yr.

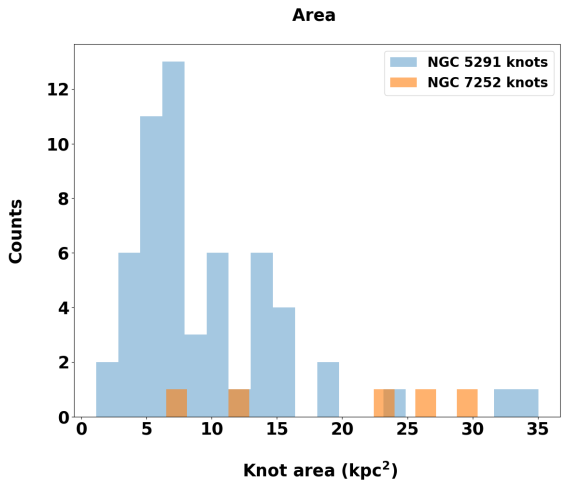

**Figure 10.** Histograms of area of the knots

## 4. Discussion
### 4.1 Comparison of NGC 7252 with NGC 5291 system

The NGC 7252 system is a notable example of a post-merger galaxy formed as a result of the merger of two spiral galaxies. The nuclei of the two parent galaxies have merged into one and the merger remnant has tidal tails, loops, and other features typical of mergers (Schweizer et al. 2013). NGC 5291 comprises of the parent galaxy which is interacting with a companion named the Seashell galaxy. The NGC 5291 system is believed to have formed as a result of a past collision

d. $\Sigma_{SFR}$ = SFR/Area of the knot

between two galaxies (Bournaud et al. 2007) and is known for its prominent ring structure in HI along which several TDGs and TDG candidates are seen.

Both NGC 7252 and NGC 5291 are examples of galaxy-galaxy interactions that exhibit unique star formation patterns. Both systems experienced a burst of star formation as a result of the collision and/or merger and host bonafide TDGs. In the NGC 7252 system, both the remnant body and the outskirts exhibit evidence of significant star formation. The central galaxies in the two systems have evolved into ETGs post-interaction. The NGC 7252 post-merger remnant shows indications of a gaseous disc and potential AGN feedback (Weaver et al. 2018; George, Joseph, Mondal, et al. 2018). Similar to this, the NGC 5291 galaxy also hosts AGN (I. Zaw, Y. .-. Chen, and G. R. Farrar 2019). These systems serve as prime examples of how galactic collisions and mergers may cause significant changes in the structure and dynamics of galaxies, resulting in bursts of star formation and the production of typical merger features. Such interacting systems are hence important laboratories to study both galaxy formation as well as evolution.

#### 4.1.1 SFR in the main body of the NGC 7252 and NGC 5291 systems

The $SFR_{FUV}(Galactic)$ in the main body of the NGC 7252 system is estimated to be $0.44 \pm 0.004$ $M_\odot$/yr. For the NGC 5291 galaxy, the $SFR_{FUV}(Galactic)$ is $0.053 \pm 0.001$ $M_\odot$/yr. The $\Sigma_{SFR}$ in the NGC 7252 main body is $(5.4 \pm 0.05) \times 10^{-3}$ $M_\odot$/yr/kpc$^2$ while for the NGC 5291 galaxy, the $\Sigma_{SFR}$ value is $(1.7 \pm 0.04) \times 10^{-3}$ $M_\odot$/yr/kpc$^2$. The main body of NGC 7252 system has a higher SFR and $\Sigma_{SFR}$ than the main body of NGC 5291 system.

#### 4.1.2 Internal attenuation and SFR of the knots

The areas of the knots in the NGC 7252 and NGC 5291 systems are given in Figure 10. Figure 11 shows the comparison of estimated parameters- β, $A_{FUV}(Internal)$, $SFR_{FUV}(uncorr)$, $SFR_{FUV}(corr)$, $\Sigma_{SFR_{FUV}}(uncorr)$ and $\Sigma_{SFR_{FUV}}(corr)$ of the knots in the NGC 7252 post merger system with that of the knots in the NGC 5291 interacting system (Rakhi et al. 2023). For the star-forming knots outside the NGC 7252 remnant, the estimated β values range from -1.66 to -1.06. For NGC 5291 system, the β range from -2.26 to -1.79. The numerical value of $A_{FUV}(Internal)$ for the knots in the NGC 7252 system ranges from 1.12 to 2.33 while for the NGC 5291 system it ranges from 0.0 to 0.87. It is observed that the SFR values (both uncorrected and corrected for extinction) of individual knots in the NGC 7252 system lie within the range of SFR values exhibited by the knots in the NGC 5291 system.

Table 3 gives a comparison of the SFR in the tidal debris and main bodies of the NGC 5291 and NGC 7252 systems. The integrated $SFR_{FUV}(corr)$ of the knots in the NGC 7252 and NGC 5291 systems are $0.46 \pm 0.03$ $M_\odot$/yr and $2.4 \pm 0.06$ $M_\odot$/yr respectively. It has to be noted that the total $SFR_{FUV}(corr)$ of the knots in the NGC 5291 system is comparable to the total SFR in the disc of normal spiral galaxies i.e., SFR > 1 $M_\odot$/yr (Boquien et al. 2009).

10    Geethika Santhosh *et al.*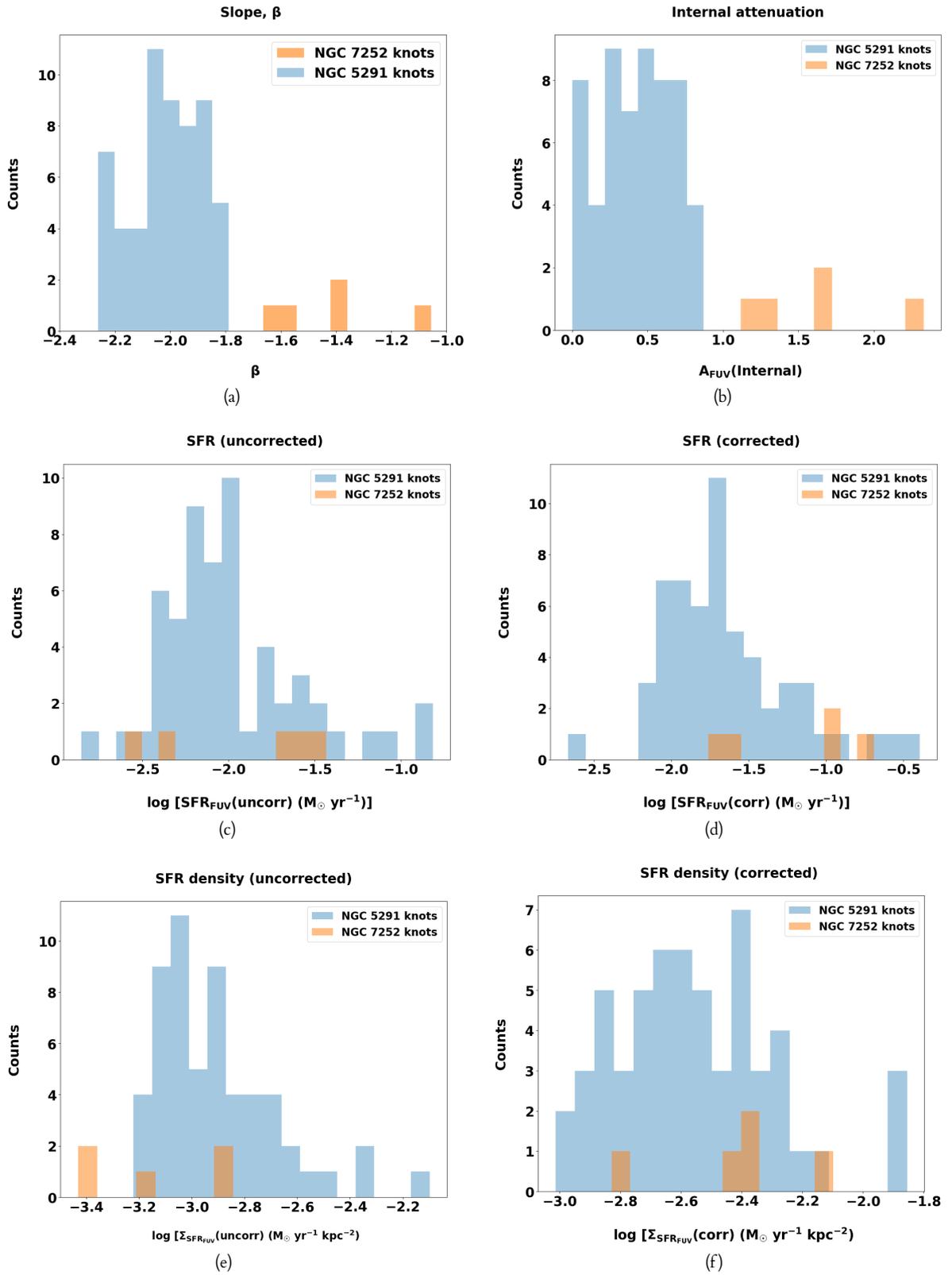

**Figure 11.** Comparison of (a) UV Slope β, (b) $A_{FUV}(Internal)$, (c) $SFR_{FUV}(uncorr)$, (d) $SFR_{FUV}(corr)$, (e) $\Sigma_{SFR_{FUV}}(uncorr)$ and (f) $\Sigma_{SFR_{FUV}}(corr)$ of the NGC 7252 post merger system with NGC 5291 interacting system.

Table 3. SFR of the knots and the main body of the interacting system NGC 5291 and post-merger system, NGC 7252

|  | Knots (including the TDGs) | | Galaxy main bodies | | |
|---|---|---|---|---|---|
|  | | | NGC 5291 | | NGC 7252 |
|  | NGC 5291 ring | NGC 7252 tails | NGC 5291 galaxy | Seashell galaxy | Post-merger remnant |
| $SFR_{FUV}(uncorr)$ ($M_\odot$/yr) | $1.0 \pm 0.005$ | $0.092 \pm 0.002$ | $0.035 \pm 0.001$ | $0.020 \pm 0.001$ | $0.36 \pm 0.003$ |
| $SFR_{FUV}$(Galactic) ($M_\odot$/yr) | $1.6 \pm 0.008$ | $0.11 \pm 0.002$ | $0.053 \pm 0.001$ | $0.030 \pm 0.001$ | $0.44 \pm 0.004$ |
| $SFR_{FUV}(corr)$ ($M_\odot$/yr) | $2.4 \pm 0.06$ | $0.46 \pm 0.03$ | - | - | - |

Table 4. Atomic, molecular and total hydrogen gas content in NGC 5291 and NGC 7252 interacting systems (Malphrus et al. 1997; Braine et al. 2001; Lelli et al. 2015)

|  | $M_{atom}$ ($M_\odot$) | $M_{mol}$ ($M_\odot$) | $M_{atom}+M_{mol}$ ($M_\odot$) |
|---|---|---|---|
| **NGC 5291** | | | |
| Main body | $\sim 1.6 \times 10^9$ | $1.7 \times 10^9$ | $\sim 3.3 \times 10^9$ |
| Collisional ring | $\sim 1.64 \times 10^{10}$ | $> 7.4 \times 10^8$ | $> 1.71 \times 10^{10}$ |
| Entire system | $1.8 \times 10^{10}$ | $> 2.44 \times 10^9$ | $> 2.04 \times 10^{10}$ |
| **NGC 7252** | | | |
| Remnant | $0.25 \times 10^9$ | $3.6 \times 10^9$ | $3.85 \times 10^9$ |
| Tidal tails | $5.2 \times 10^9$ | $> 0.02 \times 10^9$ | $> 5.22 \times 10^9$ |
| Entire system | $5.45 \times 10^9$ | $> 3.62 \times 10^9$ | $> 9.07 \times 10^9$ |

**NOTE:** The $M_{HI}$ corresponding to the NGC 7252 remnant is the mass of the so-called Western Loop (Dupraz et al. 1990; Lelli et al. 2015).

Table 5. Gas surface densities and SFR densities of TDGs of the NGC 5291 and NGC 7252 systems (Kovakkuni et al. 2023)

|  | Area ($kpc^2$) | $\Sigma_{atom}$ ($M_\odot\mathrm{pc}^{-2}$) | $\Sigma_{mol}$ ($M_\odot\mathrm{pc}^{-2}$) | $\Sigma_{atom+mol}$ ($M_\odot\mathrm{pc}^{-2}$) | $\Sigma_{SFR}$ ($M_\odot\mathrm{yr}^{-1}\mathrm{kpc}^{-2}$) | $\tau_{depletion}$ atomic (Gyr) | $\tau_{depletion}$ molecular (Gyr) | $\tau_{depletion}$ atomic + molecular (Gyr) |
|---|---|---|---|---|---|---|---|---|
| NGC 5291N | 25 | 65.6 | 2.2 | 67.8 | $0.014 \pm 0.001$ | 4.5 | 0.15 | 4.7 |
| NGC 5291S | 16 | 77.5 | 1.6 | 79.1 | $0.014 \pm 0.001$ | 5.6 | 0.12 | 5.7 |
| NGC 7252NW | 12.3 | 9.5 | 2.6 | 12.1 | $0.007 \pm 0.001$ | 1.4 | 0.39 | 1.8 |

**NOTE:** $SFR$ and $\Sigma_{SFR}$ correspond to the dust corrected values. $M_{atom}$ and $M_{mol}$ include contribution from helium and heavier elements. Note that the area corresponding to atomic gas, molecular gas, and SFR are the same and equal to the CO emitting area. The CO emitting area is given in Kovakkuni et al. (2023).

The $SFR_{FUV}(corr)$ values of the TDGs in NGC 7252 system—NGC 7252E and NGC 7252NW are $0.11 \pm 0.02$ and $0.10 \pm 0.01$ $M_\odot$/yr respectively while the same for the TDGs in NGC 5291 system—NGC 5291N, NGC 5291S and NGC5291SW are $0.40 \pm 0.03$, $0.41 \pm 0.03$ and $0.30 \pm 0.02$ $M_\odot$/yr respectively. It is observed that TDGs in the NGC 7252 system show lower SFR and $\Sigma_{SFR}$ than those in the NGC 5291 system (see also Table 5).

### 4.2 Observed trends in SFR in the NGC 7252 and NGC 5291 systems

While the main body of the NGC 7252 galaxy exhibits a higher SFR and $\Sigma_{SFR}$ than that of NGC 5291/Seashell galaxies, the NGC 7252 TDGs along the tidal tails have lower SFR and $\Sigma_{SFR}$ compared to the TDGs in the NGC 5291 collisional ring. A quantitative analysis of the gas content in the tidal debris and the main bodies of the NGC 7252 and NGC 5291 systems is essential for understanding these trends in the star formation activity. Estimates of the hydrogen gas content (atomic and molecular) is given in Table 4.

- **NGC 5291**
  Malphrus et al. 1997 performed a detailed high resolution study of the atomic hydrogen (HI) gas content in the interacting system NGC 5291 and estimated that the total HI mass of the entire system is $1.8 \times 10^{10}$ $M_\odot$. It was observed that approximately 9% of the total HI mass of the NGC 5291 system is found in its main body. The molecular hydrogen gas content in NGC 5291 main body is $1.7 \times 10^9$ $M_\odot$ (Braine et al. 2001) and the total molecular gas content in the TDGs is $7.4 \times 10^8$ $M_\odot$ (Lelli et al. 2015). Approximately, 12% of the cooler gas is in molecular form and around 70% of the total molecular gas detected in the entire system is contained in the main body of NGC 5291.
- **NGC 7252**
  For the NGC 7252 post-merger system, the mass of HI detected in the tails is $5.2 \times 10^9$ $M_\odot$ and the total HI mass of the entire system is $5.45 \times 10^9$ $M_\odot$. The remnant body is almost devoid of atomic gas (Hibbard et al. 1994). The mass of molecular hydrogen detected in the center of NGC 7252 main body is $3.6 \times 10^9$ $M_\odot$ and that detected in the tidal tails is greater than $0.02 \times 10^9$ $M_\odot$. For the entire system, the mass of molecular hydrogen detected is thus greater than $3.62 \times 10^9$ $M_\odot$ (Lelli et al. 2015). Approximately, 40% of the cooler gas is in molecular form and the main body of NGC 7252 contains almost 99% of the total molecular gas found in the whole system.

There is more gas content (atomic + molecular) in the NGC 7252 main body than in the NGC 5291 main body and less gas in the outer tidal features of NGC 7252 than in the collisional ring of NGC 5291.

#### 4.2.1 Star formation and gas content

The Kennicutt-Schmidt (KS) relation (Kennicutt 1998) is an empirical relation that connects the SFR and gas surface densities as

$$\Sigma_{SFR} \propto (\Sigma_{gas})^N$$

where N is the power law index, $\Sigma_{SFR}$ is the SFR surface density and $\Sigma_{gas} = \Sigma_{HI+H_2}$ is the total gas column density, which combines the contribution of atomic ($\Sigma_{HI}$) and molecular ($\Sigma_{H_2}$) gas.

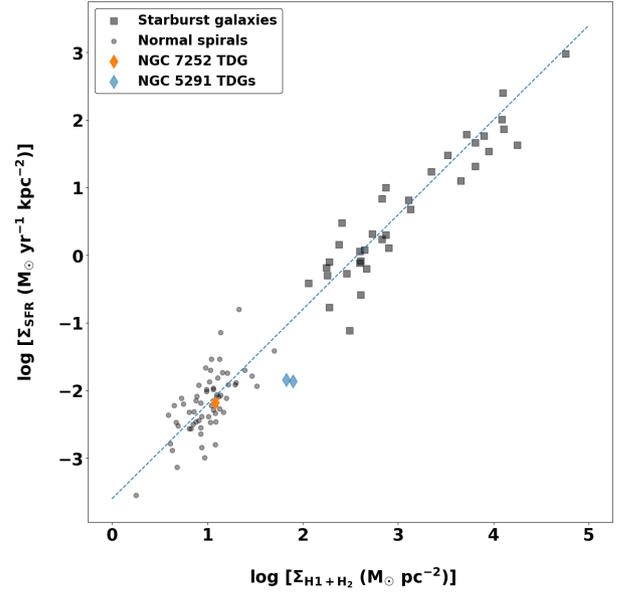

**Figure 12.** The location of TDGs on the Kennicutt-Schmidt relation (Gray markers, (Kennicutt 1998)). The blue dashed line with slope N = 1.4 is the original fit to the data (normal spirals and starbursts) from Kennicutt (1998).

The area, SFR densities, gas surface densities and the gas depletion time ($\tau_{depletion}$) for the TDGs in the NGC 5291 and NGC 7252 interacting systems are given in Table 5. The atomic and molecular masses are taken from Kovakkuni et al. (2023). High resolution CO data cubes (spatial resolution $\sim 2''$) from ALMA are used for estimating molecular mass. Molecular mass is not detected in the TDGs NGC 5291SW and NGC 7252E. The atomic mass $M_{atom}$ and the molecular mass $M_{mol}$ include contribution from helium and heavier elements. The area over which the atomic mass, molecular mass and SFR are estimated is the same and corresponds to the CO emitting area (see Kovakkuni et al. (2023)). CO is confined to a small area within the TDGs. Consequently, the total CO emitting area is smaller than the size of the TDGs observed in the UV images. As a result, the SFR estimated from the CO emitting area (to ascertain the position of the TDGs in the KS plot) will be smaller than the actual SFR of the TDGs. $\tau_{depletion}$ is estimated using the formula $\tau_{depletion} = \Sigma_{gas}/\Sigma_{SFR}$. From Table 5, it is seen that the TDGs in NGC 5291 system have higher $\Sigma_{atom+mol}$ values compared to the TDGs in NGC 7252 system. The $\Sigma_{SFR}$ values of the TDGs also exhibit the same trend.

Figure 12 shows the location of the TDGs on the KS relation. The data for normal spirals and starburst galaxies are taken from Kennicutt (1998). It is observed that the TDGs in the NGC 5291 and NGC 7252 systems are in the same regime with respect to the KS relation as normal spiral galaxies. This is in agreement with the recent results from Kovakkuni et al. (2023).

The NGC 5291 galaxy main body has a lower $\Sigma_{H_2}$ ($\Sigma_{H_2}$ = 49.7) than the NGC 7252 remnant ($\Sigma_{H_2}$ = 92.5) (Braine et al. 2001) and the $\Sigma_{SFR}$ (Galactic extinction-corrected) value of NGC 5291 galaxy is lower than that of the NGC 7252 merger remnant. The significant star formation observed in the central part of NGC 7252 may be due to the compression of gas and thus the formation of dense molecular clouds in the central part, which in turn triggered the starburst. Previous studies (Mihos and Hernquist 1994; Di Matteo et al. 2007) show that gas inflows to the central areas occur in interacting systems. Also the gas from tidal tails is falling back into the NGC 7252 merger remnant body (Hibbard et al. 1994; Hibbard and Mihos 1995; Chien and Barnes 2010). The lack of HI gas seen in the core of the NGC 7252 remnant body may be the result of a process that is changing the HI gas in the main body and in the tidal tails into other forms—molecular gas and stars. The HI gas gets compressed, condenses to molecular phase and forms stars. In the inner regions of the NGC 7252 merger remnant, our estimated SFR shows that gas is getting converted into stars at a high rate.

The observed trend in the SFR and $\Sigma_{SFR}$ of the two systems can be attributed to several factors related to the merger history and subsequent environmental circumstances. The formation scenario, the age and strength of interaction, and the current evolutionary stage of interacting systems are some of the aspects that can explain the trend. The surface density of gas in the main body and in the tidal features, as well as the quantity of cold atomic and molecular gas available for star formation in the main body and tidal features are significantly influenced by these parameters.

## 5. Conclusion

The star formation activity in the tidal tails and main body of the NGC 7252 interacting system has been investigated using FUV and NUV data from the Ultra Violet Imaging Telescope (UVIT) on board AstroSat. A comparison of the results with a study based on UVIT data on star formation in the NGC 5291 interacting system (corrected here for Galactic extinction) is also presented.

The galaxies NGC 7252 and NGC 5291 are systems that have undergone interactions and/or collisions in the past, creating unique features. Star-forming knots of young, hot stars produced by the burst of star formation that followed the collision are found within the ring structure of the NGC 5291 system. The ring structure was generated as a result of collision. Tidal tails, loops and bridges are other features of galaxies that are remnants of gravitational interactions/merger. These are composed of stars and gas pulled out during the interaction. Both the systems studied have bonafide TDGs, are located at comparable distances from the Milky Way galaxy and the same FUV and NUV filters were used for both observations. The main results of this study are summarised as follows:

1. Six star-forming regions have been identified in the NGC 7252 system including the remnant body.
2. The NGC 7252 remnant exhibits significant star formation even in the absence of internal dust correction.
3. From the UV continuum slope, β, the internal attenuation towards each of the star-forming regions has been estimated for the NGC 7252 system. The total corrected SFR of the knots in the system is $0.46 \pm 0.03$ $M_\odot$/yr. The total uncorrected SFR is estimated as $0.092 \pm 0.002$ $M_\odot$/yr.
4. A comparison of star formation in the NGC 7252 system with the star formation in the NGC 5291 system shows that the SFR of knots in NGC 7252 lie within the range of SFR observed for NGC 5291 knots.
5. For NGC 7252 system, the total SFR in the debris is $0.46 \pm 0.03$ $M_\odot/yr$ while for the NGC 5291 system it is $2.4 \pm 0.06$ $M_\odot/yr$ which is relatively high. The integrated corrected SFR of the star-forming knots along the NGC 5291 collisional ring is comparable to the total SFR in the disc of normal spiral galaxies.
6. The NGC 7252 remnant exhibits a higher SFR and SFR density than the main body of NGC 5291 while the SFR and SFR density of the NGC 7252 TDGs is lower than that of the NGC 5291 TDGs. This observed reversal in star formation trends can be attributed to the formation scenario, strength and age of interaction, and their present state of evolution. All these factors are observed to have an effect on the spatial distribution of atomic and molecular gas available for star formation. The main body of NGC 7252 has more gas and greater molecular gas surface density than the main body of NGC 5291, which results in a higher SFR as well as SFR density. The TDGs in the collisional ring of NGC 5291, in contrast, contain more gas and higher gas surface density than the TDGs in the tidal tails of NGC 7252. This is consistent with the TDGs in the NGC 5291 system having a greater SFR and SFR density than those in the NGC 7252 system.
7. The TDGs in the NGC 5291 and NGC 7252 systems fall within the same KS relation regime as normal spiral galaxies.

**Funding Statement**  The authors acknowledge the financial support of ISRO under AstroSat archival data utilisation program (No. DS_2B-13013(2)/9/2020-Sec.2). This publication uses data from the AstroSat mission of the Indian Space Research Organisation (ISRO), archived at the Indian Space Science Data Centre (ISSDC).

**Data Availability Statement**  The Astrosat UVIT imaging data underlying this article are available in ISSDC Astrobrowse archive https://astrobrowse.issdc.gov.in/astro_archive/archive/Home.jsp

**References**

Barnes, Joshua E. 1998. Dynamics of Galaxy Interactions. In *Saas-fee advanced course 26: galaxies: interactions and induced star formation,* edited by Jr. Kennicutt R. C., F. Schweizer, J. E. Barnes, D. Friedli, L. Martinet, and D. Pfenniger, 275. January.

Barnes, Joshua E., and Lars Hernquist. 1992. Formation of dwarf galaxies in tidal tails. 360, no. 6406 (December): 715–717. https://doi.org/10.1038/360715a0.

———. 1996. Transformations of Galaxies. II. Gasdynamics in Merging Disk Galaxies. 471 (November): 115. https://doi.org/10.1086/177957.

Barnes, Joshua E., and Lars E. Hernquist. 1991. Fueling Starburst Galaxies with Gas-rich Mergers. 370 (April): L65. https://doi.org/10.1086/185978.

Blumenthal, Kelly A., and Joshua E. Barnes. 2018. Go with the Flow: Understanding inflow mechanisms in galaxy collisions. 479, no. 3 (September): 3952–3965. https://doi.org/10.1093/mnras/sty1605. arXiv: 1806.05132 [astro-ph.GA].

Boquien, M., V. Buat, A. Boselli, M. Baes, G. J. Bendo, L. Ciesla, A. Cooray, et al. 2012. The IRX-β relation on subgalactic scales in star-forming galaxies of the Herschel Reference Survey. 539 (March): A145. https://doi.org/10.1051/0004-6361/201118624. arXiv: 1201.2405 [astro-ph.CO].

Boquien, M., P. -A. Duc, Y. Wu, V. Charmandaris, U. Lisenfeld, J. Braine, E. Brinks, J. Iglesias-Páramo, and C. K. Xu. 2009. Collisional Debris as Laboratories to Study Star Formation. 137, no. 6 (June): 4561–4576. https://doi.org/10.1088/0004-6256/137/6/4561. arXiv: 0903.3403 [astro-ph.CO].

Bournaud, F. 2010. Star Formation and Structure Formation in Galaxy Interactions and Mergers. In *Galaxy wars: stellar populations and star formation in interacting galaxies,* edited by B. Smith, J. Higdon, S. Higdon, and N. Bastian, 423:177. Astronomical Society of the Pacific Conference Series. June. https://doi.org/10.48550/arXiv.0909.1812. arXiv: 0909.1812 [astro-ph.CO].

Bournaud, Frédéric, Pierre-Alain Duc, Elias Brinks, Médéric Boquien, Philippe Amram, Ute Lisenfeld, Bärbel S. Koribalski, Fabian Walter, and Vassilis Charmandaris. 2007. Missing Mass in Collisional Debris from Galaxies. *Science* 316, no. 5828 (May): 1166. https://doi.org/10.1126/science.1142114. arXiv: 0705.1356 [astro-ph].

Braine, J., P. -A. Duc, U. Lisenfeld, V. Charmandaris, O. Vallejo, S. Leon, and E. Brinks. 2001. Abundant molecular gas in tidal dwarf galaxies: On-going galaxy formation. 378 (October): 51–69. https://doi.org/10.1051/0004-6361:20011109. arXiv: astro-ph/0108513 [astro-ph].

Calzetti, Daniela. 2013. Star Formation Rate Indicators. In *Secular evolution of galaxies,* edited by Jesús Falcón-Barroso and Johan H. Knapen, 419. https://doi.org/10.48550/arXiv.1208.2997.

Calzetti, Daniela, Anne L. Kinney, and Thaisa Storchi-Bergmann. 1994. Dust Extinction of the Stellar Continua in Starburst Galaxies: The Ultraviolet and Optical Extinction Law. 429 (July): 582. https://doi.org/10.1086/174346.

Cardelli, Jason A., Geoffrey C. Clayton, and John S. Mathis. 1989. The Relationship between Infrared, Optical, and Ultraviolet Extinction. 345 (October): 245. https://doi.org/10.1086/167900.

Chabrier, Gilles. 2003. Galactic Stellar and Substellar Initial Mass Function. 115, no. 809 (July): 763–795. https://doi.org/10.1086/376392. arXiv: astro-ph/0304382 [astro-ph].

Chien, L. -H., and J. E. Barnes. 2010. Dynamically driven star formation in models of NGC 7252. 407, no. 1 (September): 43–54. https://doi.org/10.1111/j.1365-2966.2010.16903.x. arXiv: 1004.3760 [astro-ph.CO].

Cortese, L., G. Gavazzi, and A. Boselli. 2008. The ultraviolet luminosity function and star formation rate of the Coma cluster. 390, no. 3 (November): 1282–1296. https://doi.org/10.1111/j.1365-2966.2008.13838.x. arXiv: 0809.0972 [astro-ph].

Dey, Arjun, David J. Schlegel, Dustin Lang, Robert Blum, Kaylan Burleigh, Xiaohui Fan, Joseph R. Findlay, et al. 2019. Overview of the DESI Legacy Imaging Surveys. 157, no. 5 (May): 168. https://doi.org/10.3847/1538-3881/ab089d. arXiv: 1804.08657 [astro-ph.IM].

Di Matteo, P., F. Combes, A. -L. Melchior, and B. Semelin. 2007. Star formation efficiency in galaxy interactions and mergers: a statistical study. 468, no. 1 (June): 61–81. https://doi.org/10.1051/0004-6361:20066959. arXiv: astro-ph/0703212 [astro-ph].

Duc, P. -A., J. Braine, U. Lisenfeld, E. Brinks, and M. Boquien. 2007. VCC 2062: an old tidal dwarf galaxy in the Virgo cluster? 475, no. 1 (November): 187–197. https://doi.org/10.1051/0004-6361:20078335. arXiv: 0709.2733 [astro-ph].

Duc, P. -A., E. Brinks, V. Springel, B. Pichardo, P. Weilbacher, and I. F. Mirabel. 2000. Formation of a Tidal Dwarf Galaxy in the Interacting System Arp 245 (NGC 2992/93). 120, no. 3 (September): 1238–1264. https://doi.org/10.1086/301516. arXiv: astro-ph/0006038 [astro-ph].

Duc, P. -A., and I. F. Mirabel. 1994. Recycled galaxies in the colliding system ARP 105. 289 (September): 83–93.

———. 1998. Young tidal dwarf galaxies around the gas-rich disturbed lenticular NGC 5291. 333 (May): 813–826.

———. 1999. Tidal Dwarf Galaxies. In *Galaxy interactions at low and high redshift,* edited by J. E. Barnes and D. B. Sanders, 186:61. January.

Dupraz, C., F. Casoli, F. Combes, and I. Kazes. 1990. CO in mergers. II. NGC 7252 : the link between mergers and ellipticals. 228 (February): L5–L8.

Elmegreen, B. G., M. Kaufman, and M. Thomasson. 1993. An Interaction Model for the Formation of Dwarf Galaxies and 10 8 M$_{sun}$ Clouds in Spiral Disks. 412 (July): 90. https://doi.org/10.1086/172903.

George, K., P. Joseph, P. Côté, S. K. Ghosh, J. B. Hutchings, R. Mohan, J. Postma, et al. 2018. Dissecting star formation in the "Atoms-for-Peace" galaxy. UVIT observations of the post-merger galaxy NGC7252. 614 (June): A130. https://doi.org/10.1051/0004-6361/201832705. arXiv: 1802.09493 [astro-ph.GA].

George, K., P. Joseph, C. Mondal, A. Devaraj, A. Subramaniam, C. S. Stalin, P. Côté, et al. 2018. UVIT observations of the star-forming ring in NGC 7252: Evidence of possible AGN feedback suppressing central star formation. 613 (May): L9. https://doi.org/10.1051/0004-6361/201833232. arXiv: 1805.03543 [astro-ph.GA].

George, Koshy. 2023. Decoding NGC 7252 as a blue elliptical galaxy. 671 (March): A166. https://doi.org/10.1051/0004-6361/202345837. arXiv: 2302.03369 [astro-ph.GA].

Hancock, Mark, Beverly J. Smith, Curtis Struck, Mark L. Giroux, Philip N. Appleton, Vassilis Charmandaris, and William T. Reach. 2007. Large-Scale Star Formation Triggering in the Low-Mass Arp 82 System: A Nearby Example of Galaxy Downsizing Based on UV/Optical/Mid-IR Imaging. 133, no. 2 (February): 676–693. https://doi.org/10.1086/510241. arXiv: astro-ph/0610421 [astro-ph].

Hancock, Mark, Beverly J. Smith, Curtis Struck, Mark L. Giroux, and Sabrina Hurlock. 2009. Candidate Tidal Dwarf Galaxies in Arp 305: Lessons on Dwarf Detachment and Globular Cluster Formation. 137, no. 6 (June): 4643–4654. https://doi.org/10.1088/0004-6256/137/6/4643. arXiv: 0904.0670 [astro-ph.GA].

Hao, Cai-Na, Robert C. Kennicutt, Benjamin D. Johnson, Daniela Calzetti, Daniel A. Dale, and John Moustakas. 2011. Dust-corrected Star Formation Rates of Galaxies. II. Combinations of Ultraviolet and Infrared Tracers. 741, no. 2 (November): 124. https://doi.org/10.1088/0004-637X/741/2/124. arXiv: 1108.2837 [astro-ph.CO].

Hibbard, J. E., Puragra Guhathakurta, J. H. van Gorkom, and Francois Schweizer. 1994. Cold, Warm, and Hot Gas in the Late-Stage Merger NGC 7252. 107 (January): 67. https://doi.org/10.1086/116835.

Hibbard, J. E., and J. Christopher Mihos. 1995. Dynamical Modeling of NGC 7252 and the Return of Tidal Material. 110 (July): 140. https://doi.org/10.1086/117502. arXiv: astro-ph/9503030 [astro-ph].

Hopkins, Philip F., Lars Hernquist, Thomas J. Cox, and Dušan Kereš. 2008. A Cosmological Framework for the Co-Evolution of Quasars, Supermassive Black Holes, and Elliptical Galaxies. I. Galaxy Mergers and Quasar Activity. 175, no. 2 (April): 356–389. https://doi.org/10.1086/524362. arXiv: 0706.1243 [astro-ph].

Hota, Ananda, Ashish Devaraj, Ananta C. Pradhan, C. S. Stalin, Koshy George, Abhisek Mohapatra, Soo-Chang Rey, et al. 2021. The sharpest ultraviolet view of the star formation in an extreme environment of the nearest Jellyfish Galaxy IC 3418. *Journal of Astrophysics and Astronomy* 42, no. 2 (October): 86. https://doi.org/10.1007/s12036-021-09764-w. arXiv: 2104.14325 [astro-ph.GA].

Hunsberger, Sally D., Jane C. Charlton, and Dennis Zaritsky. 1996. The Formation of Dwarf Galaxies in Tidal Debris: A Study of the Compact Group Environment. 462 (May): 50. https://doi.org/10.1086/177126. arXiv: astro-ph/9510160 [astro-ph].

Hunter, Deidre A., Sally D. Hunsberger, and Erin W. Roye. 2000. Identifying Old Tidal Dwarf Irregulars. 542, no. 1 (October): 137–142. https://doi.org/10.1086/309542. arXiv: astro-ph/0005257 [astro-ph].

Iglesias-Páramo, J., V. Buat, T. T. Takeuchi, K. Xu, S. Boissier, A. Boselli, D. Burgarella, et al. 2006. Star Formation in the Nearby Universe: The Ultraviolet and Infrared Points of View. 164, no. 1 (May): 38–51. https://doi.org/10.1086/502628. arXiv: astro-ph/0601235 [astro-ph].

Joseph, Prajwel, P. Sreekumar, C. S. Stalin, K. T. Paul, Chayan Mondal, Koshy George, and Blesson Mathew. 2022. UVIT view of Centaurus A: a detailed study on positive AGN feedback. 516, no. 2 (October): 2300–2313. https://doi.org/10.1093/mnras/stac2388. arXiv: 2208.10209 [astro-ph.GA].

Kaviraj, S., K. Schawinski, J. E. G. Devriendt, I. Ferreras, S. Khochfar, S. -J. Yoon, S. K. Yi, et al. 2007. UV-Optical Colors As Probes of Early-Type Galaxy Evolution. 173, no. 2 (December): 619–642. https://doi.org/10.1086/516633. arXiv: astro-ph/0601029 [astro-ph].

Kaviraj, Sugata, Daniel Darg, Chris Lintott, Kevin Schawinski, and Joseph Silk. 2012. Tidal dwarf galaxies in the nearby Universe. 419, no. 1 (January): 70–79. https://doi.org/10.1111/j.1365-2966.2011.19673.x. arXiv: 1108.4410 [astro-ph.CO].

Kaviraj, Sugata, Kok-Meng Tan, Richard S. Ellis, and Joseph Silk. 2011. A coincidence of disturbed morphology and blue UV colour: minor-merger-driven star formation in early-type galaxies at z∼ 0.6. 411, no. 4 (March): 2148–2160. https://doi.org/10.1111/j.1365-2966.2010.17754.x. arXiv: 1001.2141 [astro-ph.CO].

Kennicutt, Jr., Robert C. 1998. The Global Schmidt Law in Star-forming Galaxies. 498, no. 2 (May): 541–552. https://doi.org/10.1086/305588. arXiv: astro-ph/9712213 [astro-ph].

Kennicutt, Robert C., and Neal J. Evans. 2012. Star Formation in the Milky Way and Nearby Galaxies. 50 (September): 531–608. https://doi.org/10.1146/annurev-astro-081811-125610. arXiv: 1204.3552 [astro-ph.GA].

Komatsu, E., K. M. Smith, J. Dunkley, C. L. Bennett, B. Gold, G. Hinshaw, N. Jarosik, et al. 2011. Seven-year Wilkinson Microwave Anisotropy Probe (WMAP) Observations: Cosmological Interpretation. 192, no. 2 (February): 18. https://doi.org/10.1088/0067-0049/192/2/18. arXiv: 1001.4538 [astro-ph.CO].

Kovakkuni, N., F. Lelli, P. -A. Duc, M. Boquien, J. Braine, E. Brinks, V. Charmandaris, et al. 2023. Molecular and ionized gas in tidal dwarf galaxies: the spatially resolved star formation relation. 526, no. 2 (December): 1940–1950. https://doi.org/10.1093/mnras/stad2790. arXiv: 2309.06478 [astro-ph.GA].

Kroupa, Pavel. 2001. On the variation of the initial mass function. 322, no. 2 (April): 231–246. https://doi.org/10.1046/j.1365-8711.2001.04022.x. arXiv: astro-ph/0009005 [astro-ph].

Larson, Richard B. 1990. Galaxy Building. 102 (July): 709. https://doi.org/10.1086/132694.

Lelli, Federico, Pierre-Alain Duc, Elias Brinks, Frédéric Bournaud, Stacy S. McGaugh, Ute Lisenfeld, Peter M. Weilbacher, et al. 2015. Gas dynamics in tidal dwarf galaxies: Disc formation at z = 0. 584 (December): A113. https://doi.org/10.1051/0004-6361/201526613. arXiv: 1509.05404 [astro-ph.GA].

Li, Wenhao, Preethi Nair, Jimmy Irwin, Sara Ellison, Shobita Satyapal, Niv Drory, Amy Jones, et al. 2023. A Multiwavelength Study of Active Galactic Nuclei in Post-merger Remnants. 944, no. 2 (February): 168. https://doi.org/10.3847/1538-4357/acb13d. arXiv: 2301.06186 [astro-ph.GA].

Longmore, A. J., T. G. Hawarden, R. D. Cannon, D. A. Allen, U. Mebold, W. M. Goss, and K. Reif. 1979. NGC 5291: a massive, gas-rich and highly peculiar lenticular in the IC 4329 cluster. 188 (July): 285–296. https://doi.org/10.1093/mnras/188.2.285.

Mahajan, Smriti, Kulinder Pal Singh, Joseph E. Postma, Kala G. Pradeep, Koshy George, and Patrick Côté. 2022. Deepest far ultraviolet view of a central field in the Coma cluster by AstroSat UVIT. 39 (October): e048. https://doi.org/10.1017/pasa.2022.45. arXiv: 2209.05886 [astro-ph.GA].

Malphrus, Benjamin K., Caroline E. Simpson, S. T. Gottesman, and Timothy G. Hawarden. 1997. NGC 5291: implications for the formation of dwarf galaxies. 114 (October): 1427–1446. https://doi.org/10.1086/118574.

Meurer, Gerhardt R., Timothy M. Heckman, and Daniela Calzetti. 1999. Dust Absorption and the Ultraviolet Luminosity Density at z ∼3 as Calibrated by Local Starburst Galaxies. 521, no. 1 (August): 64–80. https://doi.org/10.1086/307523. arXiv: astro-ph/9903054 [astro-ph].

Mihos, J. Christopher, and Lars Hernquist. 1994. Triggering of Starbursts in Galaxies by Minor Mergers. 425 (April): L13. https://doi.org/10.1086/187299.

———. 1996. Gasdynamics and Starbursts in Major Mergers. 464 (June): 641. https://doi.org/10.1086/177353. arXiv: astro-ph/9512099 [astro-ph].

Mirabel, I. F., H. Dottori, and D. Lutz. 1992. Genesis of a dwarf galaxy from the debris of the Antennae. 256 (March): L19–L22.

Mirabel, I. F., D. Lutz, and J. Maza. 1991. The Superantennae. 243 (March): 367.

Mondal, Chayan, Annapurni Subramaniam, and Koshy George. 2018. UVIT Imaging of WLM: Demographics of Star-forming Regions in the Nearby Dwarf Irregular Galaxy. 156, no. 3 (September): 109. https://doi.org/10.3847/1538-3881/aad4f6. arXiv: 1807.07359 [astro-ph.GA].

———. 2019. Ultraviolet Imaging Telescope View of Dwarf Irregular Galaxy IC 2574: Is the Star Formation Triggered Due to Expanding H I Shells? 158, no. 6 (December): 229. https://doi.org/10.3847/1538-3881/ab4ea1. arXiv: 1906.10660 [astro-ph.GA].

———. 2021. A tale of two nearby dwarf irregular galaxies WLM and IC 2574: As revealed by UVIT. *Journal of Astrophysics and Astronomy* 42, no. 2 (October): 50. https://doi.org/10.1007/s12036-021-09761-z. arXiv: 2105.13048 [astro-ph.GA].

Mondal, Chayan, Annapurni Subramaniam, Koshy George, Joseph E. Postma, Smitha Subramanian, and Sudhanshu Barway. 2021. Tracing Young Star-forming Clumps in the Nearby Flocculent Spiral Galaxy NGC 7793 with UVIT Imaging. 909, no. 2 (March): 203. https://doi.org/10.3847/1538-4357/abe0b4. arXiv: 2101.11314 [astro-ph.GA].

O'Donnell, James E. 1994. R v-dependent Optical and Near-Ultraviolet Extinction. 422 (February): 158. https://doi.org/10.1086/173713.

Overzier, Roderik A., Timothy M. Heckman, Jing Wang, Lee Armus, Veronique Buat, Justin Howell, Gerhardt Meurer, et al. 2011. Dust Attenuation in UV-selected Starbursts at High Redshift and Their Local Counterparts: Implications for the Cosmic Star Formation Rate Density. 726, no. 1 (January): L7. https://doi.org/10.1088/2041-8205/726/1/L7. arXiv: 1011.6098 [astro-ph.CO].

Poggianti, Bianca M., Alessandro Ignesti, Myriam Gitti, Anna Wolter, Fabrizio Brighenti, Andrea Biviano, Koshy George, et al. 2019. GASP XXIII: A Jellyfish Galaxy as an Astrophysical Laboratory of the Baryonic Cycle. 887, no. 2 (December): 155. https://doi.org/10.3847/1538-4357/ab5224. arXiv: 1910.11622 [astro-ph.GA].

Postma, Joseph E., and Denis Leahy. 2017. CCDLAB: A Graphical User Interface FITS Image Data Reducer, Viewer, and Canadian UVIT Data Pipeline. 129, no. 981 (November): 115002. https://doi.org/10.1088/1538-3873/aa8800.

———. 2020. An Algorithm for Coordinate Matching in World Coordinate Solutions. 132, no. 1011 (May): 054503. https://doi.org/10.1088/1538-3873/ab7ee8.

———. 2021. UVIT data reduction pipeline: A CCDLAB and UVIT tutorial. *Journal of Astrophysics and Astronomy* 42, no. 2 (October): 30. https://doi.org/10.1007/s12036-020-09689-w.

Postma, Joseph E., Shyam N. Tandon, Denis Leahy, and John Hutchings. 2023. Assessing AstroSat bus jitter from UVIT data of small magellenic cloud. *Journal of Astrophysics and Astronomy* 44, no. 1 (June): 12. https://doi.org/10.1007/s12036-022-09901-z.

Rakhi, R., Geethika Santhosh, Prajwel Joseph, Koshy George, Smitha Subramanian, Indulekha Kavila, J. Postma, et al. 2023. UVIT view of NGC 5291: Ongoing star formation in tidal dwarf galaxies at 0.35 kpc resolution. 522, no. 1 (June): 1196–1207. https://doi.org/10.1093/mnras/stad970. arXiv: 2304.07244 [astro-ph.GA].

Robin, T., Sreeja S. Kartha, R. Akhil Krishna, Ujjwal Krishnan, Blesson Mathew, T. B. Cysil, Narendra Nath Patra, and B. Shridharan. 2024. The Interaction Jigsaw: investigating star formation in interacting galaxies. 534, no. 3 (November): 1902–1912. https://doi.org/10.1093/mnras/stae2211. arXiv: 2409.15497 [astro-ph.GA].

Robotham, A. S. G., L. J. M. Davies, S. P. Driver, S. Koushan, D. S. Taranu, S. Casura, and J. Liske. 2018. ProFound: Source Extraction and Application to Modern Survey Data. 476, no. 3 (May): 3137–3159. https://doi.org/10.1093/mnras/sty440. arXiv: 1802.00937 [astro-ph.IM].

Salpeter, Edwin E. 1955. The Luminosity Function and Stellar Evolution. 121 (January): 161. https://doi.org/10.1086/145971.

Schawinski, K., S. Kaviraj, S. Khochfar, S. -J. Yoon, S. K. Yi, J. -M. Deharveng, A. Boselli, et al. 2007. The Effect of Environment on the Ultraviolet Color-Magnitude Relation of Early-Type Galaxies. 173, no. 2 (December): 512–523. https://doi.org/10.1086/516631. arXiv: astro-ph/0601036 [astro-ph].

Schawinski, Kevin, Nathan Dowlin, Daniel Thomas, C. Megan Urry, and Edward Edmondson. 2010. The Role of Mergers in Early-type Galaxy Evolution and Black Hole Growth. 714, no. 1 (May): L108–L112. https://doi.org/10.1088/2041-8205/714/1/L108. arXiv: 1003.4018 [astro-ph.CO].

Schechtman-Rook, Andrew, and Kelley M. Hess. 2012. NGC 4656UV: A UV-selected Tidal Dwarf Galaxy Candidate. 750, no. 2 (May): 171. https://doi.org/10.1088/0004-637X/750/2/171. arXiv: 1203.1319 [astro-ph.GA].

Schlafly, Edward F., and Douglas P. Finkbeiner. 2011. Measuring Reddening with Sloan Digital Sky Survey Stellar Spectra and Recalibrating SFD. 737, no. 2 (August): 103. https://doi.org/10.1088/0004-637X/737/2/103. arXiv: 1012.4804 [astro-ph.GA].

Schweizer, F. 1982. Colliding and merging galaxies. I. Evidence for the recent merging of two disk galaxies in NGC 7252. 252 (January): 455–460. https://doi.org/10.1086/159573.

Schweizer, Francois, and Patrick Seitzer. 1992. Correlation Between UBV Colors and Fine Structure in E and S0. 104 (September): 1039. https://doi.org/10.1086/116296.

Schweizer, François, Patrick Seitzer, Daniel D. Kelson, Edward V. Villanueva, and Gregory L. Walth. 2013. The [O III] Nebula of the Merger Remnant NGC 7252: A Likely Faint Ionization Echo. 773, no. 2 (August): 148. https://doi.org/10.1088/0004-637X/773/2/148. arXiv: 1307.2233 [astro-ph.CO].

Sengupta, Chandreyee, T. C. Scott, K. S. Dwarakanath, D. J. Saikia, and B. W. Sohn. 2014. H I in the Arp 202 system and its tidal dwarf candidate. 444, no. 1 (October): 558–565. https://doi.org/10.1093/mnras/stu1463. arXiv: 1407.7643 [astro-ph.GA].

Sengupta, Chandreyee, T. C. Scott, S. Paudel, K. S. Dwarakanath, D. J. Saikia, and B. W. Sohn. 2017. H I, star formation and tidal dwarf candidate in the Arp 305 system. 469, no. 3 (August): 3629–3640. https://doi.org/10.1093/mnras/stx885. arXiv: 1704.04344 [astro-ph.GA].

Tandon, S. N., J. Postma, P. Joseph, A. Devaraj, A. Subramaniam, I. V. Barve, K. George, et al. 2020. Additional Calibration of the Ultraviolet Imaging Telescope on Board AstroSat. 159, no. 4 (April): 158. https://doi.org/10.3847/1538-3881/ab72a3. arXiv: 2002.01159 [astro-ph.IM].

Tandon, S. N., Annapurni Subramaniam, V. Girish, J. Postma, K. Sankarasubramanian, S. Sriram, C. S. Stalin, et al. 2017. In-orbit Calibrations of the Ultraviolet Imaging Telescope. 154, no. 3 (September): 128. https://doi.org/10.3847/1538-3881/aa8451. arXiv: 1705.03715 [astro-ph.IM].

Toomre, Alar, and Juri Toomre. 1972. Galactic Bridges and Tails. 178 (December): 623–666. https://doi.org/10.1086/151823.

Ujjwal, K., Sreeja S. Kartha, Smitha Subramanian, Koshy George, Robin Thomas, and Blesson Mathew. 2022. Understanding the secular evolution of NGC 628 using UltraViolet Imaging Telescope. 516, no. 2 (October): 2171–2180. https://doi.org/10.1093/mnras/stac2285. arXiv: 2208.05999 [astro-ph.GA].

Weaver, J., B. Husemann, K. Kuntschner, I. Martín-Navarro, F. Bournaud, P. -A. Duc, E. Emsellem, D. Krajnović, M. Lyubenova, and R. M. McDermid. 2018. History and destiny of an emerging early-type galaxy. New IFU insights on the major-merger remnant NGC 7252. 614 (June): A32. https://doi.org/10.1051/0004-6361/201732448. arXiv: 1801.09691 [astro-ph.GA].

Yoon, Yongmin, Changbom Park, Haeun Chung, and Richard R. Lane. 2022. Evidence for Impact of Galaxy Mergers on Stellar Kinematics of Early-type Galaxies. 925, no. 2 (February): 168. https://doi.org/10.3847/1538-4357/ac415d. arXiv: 2112.13703 [astro-ph.GA].

Yoshida, Michitoshi, Yoshiaki Taniguchi, and Takashi Murayama. 1994. A Forming Dwarf Galaxy in a Tidal Tail of the Merging Galaxy NGC 2782. 46 (December): L195–L198.

Zaw, I., Y. -P. Chen, and G. R. Farrar. 2019. VizieR Online Data Catalog: A uniformly selected, all-sky, optical AGN catalog (Zaw+, 2019). *VizieR Online Data Catalog* (February): J/ApJ/872/134.

Zaw, Ingyin, Yan-Ping Chen, and Glennys R. Farrar. 2019. A Uniformly Selected, All-sky, Optical AGN Catalog. 872, no. 2 (February): 134. https://doi.org/10.3847/1538-4357/aaffaf. arXiv: 1902.03799 [astro-ph.GA].

Zwicky, F. 1956. Multiple Galaxies. *Ergebnisse der exakten Naturwissenschaften* 29 (January): 344–385.

**Table 6.** Observed FUV, NUV and DECaLS r band fluxes for the star forming regions in NGC 5291 system (raw fluxes uncorrected for Galactic extinction and internal attenuation).

| No. | Region | RA (Degree) | DEC (Degree) | Knot area ($kpc^2$) | Flux FUV ($erg/s/cm^2$/Å) | eFlux FUV ($erg/s/cm^2$/Å) | Flux NUV ($erg/s/cm^2$/Å) | eFlux NUV ($erg/s/cm^2$/Å) | Flux r (nanomaggy) |
|---|---|---|---|---|---|---|---|---|---|
| 1 | D | 335.1866 | -24.6783 | 81.8 | 5.51e-15 | 4.57e-17 | 4.74e-15 | 1.15e-17 | 8221.27 |
| 2 | E | 335.1728 | -24.6793 | 25.61 | 5.57e-16 | 1.45e-17 | 2.84e-16 | 2.82e-18 | 182.65 |
| 3 | H | 335.14 | -24.6228 | 22.68 | 4.42e-16 | 1.3e-17 | 1.98e-16 | 2.36e-18 | 72.75 |
| 4 | B | 335.2332 | -24.6859 | 30.37 | 2.89e-16 | 1.05e-17 | 1.49e-16 | 2.04e-18 | 33.43 |
| 5 | A | 335.2473 | -24.6863 | 11.68 | 6.54e-17 | 4.98e-18 | 3.06e-17 | 9.26e-19 | 4.68 |
| 6 | C | 335.2077 | -24.6835 | 6.5 | 3.79e-17 | 3.79e-18 | 2.29e-17 | 8.01e-19 | 10.98 |

**NOTE:** Effective wavelengths of the filters: UVIT FUV: 1481 Å , UVIT NUV: 2418 Å, DECaLS r-band: 6382.6 Å

**Table 7.** Observed FUV, NUV and DECaLS r band fluxes for the star forming regions in NGC 5291 system (raw fluxes uncorrected for Galactic extinction and internal attenuation). The FUV and NUV fluxes are the same as given in the Appendix of (Rakhi et al. 2023).

| No. | Region Rakhi et al. (2023) | RA (Degree) | DEC (Degree) | Knot area ($kpc^2$) | Flux FUV ($erg/s/cm^2$/Å) | eFlux FUV ($erg/s/cm^2$/Å) | Flux NUV ($erg/s/cm^2$/Å) | eFlux NUV ($erg/s/cm^2$/Å) | Flux r (nanomaggy) |
|---|---|---|---|---|---|---|---|---|---|
| 1 | 1 | 206.7495 | -30.4242 | 7.01 | 1.07e-16 | 6.32e-18 | 3.66e-17 | 1e-18 | 3.47 |
| 2 | 2 | 206.7513 | -30.4195 | 10.76 | 1.28e-1 | 6.92e-18 | 4.69e-17 | 1.14e-18 | 6.18 |
| 3 | 3 | 206.7523 | -30.4221 | 9.71 | 4e-16 | 1.22e-17 | 1.5e-16 | 2.03e-18 | 16.48 |
| 4 | 4 | 206.7543 | -30.4238 | 13.7 | 1.95e-16 | 8.56e-18 | 7.01e-17 | 1.39e-18 | 13.11 |
| 5 | 5 | 206.7647 | -30.3707 | 11.52 | 7.11e-16 | 1.63e-17 | 2.64e-16 | 2.7e-18 | 27.18 |
| 6 | 6 | 206.7928 | -30.4638 | 6.93 | 1.61e-16 | 7.76e-18 | 6.22e-17 | 1.31e-18 | 5.61 |
| 7 | 7 | 206.8064 | -30.3829 | 7.62 | 1.09e-16 | 6.4e-18 | 4.09e-17 | 1.06e-18 | 4.09 |
| 8 | 8 | 206.8194 | -30.4738 | 14.28 | 3.39e-16 | 1.13e-17 | 1.22e-16 | 1.83e-18 | 18.47 |
| 9 | 9 | 206.8217 | -30.3431 | 5.25 | 1.9e-16 | 8.44e-18 | 7.66e-17 | 1.45e-18 | 6.75 |
| 10 | 10 | 206.8228 | -30.4483 | 6.69 | 1.05e-16 | 6.28e-18 | 3.65e-17 | 1e-18 | 5.78 |
| 11 | 11 | 206.8234 | -30.4719 | 19.15 | 2.78e-15 | 3.23e-17 | 9.98e-16 | 5.24e-18 | 96.41 |
| 12 | 12 | 206.8236 | -30.3338 | 13.54 | 3.24e-16 | 1.1e-17 | 1.37e-16 | 1.94e-18 | 19.93 |
| 13 | 13 | 206.8237 | -30.3379 | 2.61 | 6.31e-17 | 4.87e-18 | 2.58e-17 | 8.42e-19 | 2.96 |
| 14 | 14 | 206.8243 | -30.4426 | 7.16 | 1.21e-16 | 6.74e-18 | 4.52e-17 | 1.11e-18 | 9.07 |
| 15 | 15 | 206.8245 | -30.3662 | 10.89 | 1.86e-16 | 8.34e-18 | 7.73e-17 | 1.46e-18 | 8.25 |
| 16 | 16 | 206.8252 | -30.3528 | 9.77 | 4.61e-16 | 1.31e-17 | 1.81e-16 | 2.23e-18 | 22.97 |
| 17 | 17 | 206.8252 | -30.3428 | 7.71 | 1.88e-16 | 8.39e-18 | 7.01e-17 | 1.39e-18 | 8.04 |
| 18 | 18 | 206.8255 | -30.3642 | 7.32 | 2.06e-16 | 8.79e-18 | 8.21e-17 | 1.5e-18 | 10.52 |
| 19 | 19 | 206.8259 | -30.3583 | 9.05 | 1.34e-16 | 7.09e-18 | 5.49e-17 | 1.23e-18 | 7.03 |
| 20 | 20 | 206.8271 | -30.344 | 7.6 | 2.03e-16 | 8.73e-18 | 8.7e-17 | 1.55e-18 | 11.67 |
| 21 | 21 | 206.8276 | -30.3544 | 5.37 | 1.49e-16 | 7.48e-18 | 5.69e-17 | 1.25e-18 | 7.06 |
| 22 | 22 | 206.8276 | -30.3571 | 3.93 | 7.96e-17 | 5.46e-18 | 3.21e-17 | 9.39e-19 | 4.52 |
| 23 | 23 | 206.8315 | -30.4481 | 14.97 | 5.32e-16 | 1.41e-17 | 2.18e-16 | 2.45e-18 | 21.86 |
| 23 | 24 | 206.8329 | -30.3398 | 24.21 | 6.13e-16 | 1.52e-17 | 2.42e-16 | 2.58e-18 | 31.19 |
| 25 | 25 | 206.8355 | -30.4565 | 13.59 | 1.66e-16 | 7.88e-18 | 6.15e-17 | 1.3e-18 | 9.38 |
| 26 | 26 | 206.8355 | -30.3479 | 33.32 | 2.59e-15 | 3.12e-17 | 1.06e-15 | 5.39e-18 | 150.82 |
| 27 | 27 | 206.8366 | -30.4458 | 16.12 | 2.68e-16 | 1e-17 | 9.43e-17 | 1.61e-18 | 8.47 |
| 28 | 28 | 206.837 | -30.3589 | 7.04 | 1.4e-16 | 7.25e-18 | 5.38e-17 | 1.22e-18 | 8.02 |
| 29 | 29 | 206.8383 | -30.3397 | 5.95 | 1.06e-16 | 6.3e-18 | 3.97e-17 | 1.04e-18 | 3.32 |
| 30 | 30 | 206.8386 | -30.3574 | 6.0 | 8.77e-17 | 5.73e-18 | 3.22e-17 | 9.41e-19 | 5.03 |
| 31 | 31 | 206.8388 | -30.3506 | 9.3 | 1.37e-16 | 7.17e-18 | 5.43e-17 | 1.22e-18 | 10.02 |
| 32 | 32 | 206.8408 | -30.345 | 7.75 | 1.7e-16 | 7.98e-18 | 6.7e-17 | 1.36e-18 | 7.01 |
| 33 | 33 | 206.8425 | -30.383 | 4.59 | 7.61e-17 | 5.34e-18 | 2.87e-17 | 8.87e-19 | 4.94 |
| 34 | 34 | 206.8433 | -30.3493 | 5.25 | 7.27e-17 | 5.22e-18 | 2.52e-17 | 8.33e-19 | 4.06 |
| 35 | 35 | 206.844 | -30.3698 | 4.95 | 7.92e-17 | 5.45e-18 | 2.74e-17 | 8.68e-19 | 2.57 |
| 36 | 36 | 206.845 | -30.4353 | 7.79 | 8.52e-17 | 5.65e-18 | 3.58e-17 | 9.91e-19 | 9.26 |
| 37 | 37 | 206.8454 | -30.4647 | 35.04 | 1.25e-15 | 2.17e-17 | 4.66e-16 | 3.58e-18 | 52.62 |
| 38 | 38 | 206.8457 | -30.364 | 3.11 | 4.95e-17 | 4.31e-18 | 2.01e-17 | 7.44e-19 | 2.36 |
| 39 | 39 | 206.8459 | -30.4583 | 19.23 | 1.55e-15 | 2.41e-17 | 6.29e-16 | 4.16e-18 | 72.84 |
| 40 | 40 | 206.8462 | -30.4619 | 1.13 | 2.54e-17 | 3.08e-18 | 8.8e-18 | 4.92e-19 | 1.76 |
| 41 | 41 | 206.8477 | -30.4558 | 6.98 | 9.9e-17 | 6.09e-18 | 3.78e-17 | 1.02e-18 | 3.65 |
| 42 | 42 | 206.8479 | -30.4496 | 7.48 | 2.29e-16 | 9.27e-18 | 8.23e-17 | 1.5e-18 | 8.64 |
| 43 | 43 | 206.8482 | -30.4475 | 4.05 | 1.87e-16 | 8.36e-18 | 7.01e-17 | 1.39e-18 | 7.5 |
| 44 | 44 | 206.8488 | -30.4452 | 14.14 | 2.85e-16 | 1.03e-17 | 1.1e-16 | 1.74e-18 | 11.37 |
| 45 | 45 | 206.8489 | -30.4625 | 5.8 | 1.71e-16 | 8e-18 | 6.65e-17 | 1.35e-18 | 8.69 |
| 46 | 46 | 206.8493 | -30.4521 | 15.51 | 5.25e-16 | 1.4e-17 | 2.06e-16 | 2.38e-18 | 23.31 |
| 47 | 47 | 206.8501 | -30.4391 | 9.18 | 2.7e-16 | 1.01e-17 | 1.09e-16 | 1.73e-18 | 17.53 |
| 48 | 48 | 206.8509 | -30.4322 | 3.88 | 8.61e-17 | 5.68e-18 | 3.65e-17 | 1e-18 | 6.22 |
| 49 | 49 | 206.8514 | -30.4342 | 4.4 | 7.78e-17 | 5.4e-18 | 3.04e-17 | 9.14e-19 | 5.61 |
| 50 | 50 | 206.8518 | -30.443 | 13.26 | 1.96e-16 | 8.56e-18 | 7.75e-17 | 1.46e-18 | 11.78 |
| 51 | 51 | 206.8576 | -30.4298 | 5.94 | 1.22e-16 | 6.77e-18 | 4.78e-17 | 1.15e-18 | 12.55 |
| 52 | 52 | 206.8583 | -30.4277 | 5.59 | 1.23e-16 | 6.8e-18 | 5.04e-17 | 1.18e-18 | 12.43 |
| 53 | 53 | 206.8606 | -30.4306 | 15.54 | 5.45e-16 | 1.43e-17 | 2.02e-16 | 2.35e-18 | 43.41 |
| 54 | 54 | 206.8663 | -30.4148 | 4.48 | 7.02e-17 | 5.13e-18 | 2.83e-17 | 8.82e-19 | 11.33 |
| 55 | 55 | 206.7629 | -30.4293 | 9.88 | 1.11e-16 | 6.46e-18 | 3.88e-17 | 1.03e-18 | 6.42 |
| 56 | 56 | 206.7918 | -30.3138 | 5.37 | 8.8e-17 | 5.74e-18 | 3.63e-17 | 9.99e-19 | 11.02 |
| 57 | 57 | 206.8662 | -30.4428 | 10.86 | 1.41e-16 | 7.27e-18 | 4.79e-17 | 1.15e-18 | 13.62 |
| 58 | NGC 5291 | 206.8514 | -30.4062 | 20.3 | 4.35e-16 | 1.28e-17 | 4.3e-16 | 3.44e-18 | 3876.61 |
| 59 | NGC 5291 | 206.8547 | -30.4066 | 5.84 | 9.77e-17 | 6.05e-18 | 7.12e-17 | 1.4e-18 | 381.95 |
| 60 | NGC 5291 | 206.8502 | -30.4107 | 2.8 | 4.91e-17 | 4.29e-18 | 2.85e-17 | 8.85e-19 | 137.87 |
| 61 | NGC 5291 | 206.8513 | -30.4039 | 2.78 | 5.49e-17 | 4.54e-18 | 3.8e-17 | 1.02e-18 | 192.53 |
| 62 | Seashell | 206.8471 | -30.4175 | 14.47 | 2.68e-16 | 1e-17 | 2.78e-16 | 2.76e-18 | 1674.84 |
| 63 | Seashell | 206.8503 | -30.4159 | 2.81 | 5.82e-17 | 4.67e-18 | 5.28e-17 | 1.2e-18 | 233.28 |
| 64 | Seashell | 206.8493 | -30.4164 | 1.07 | 2.85e-17 | 3.27e-18 | 2.25e-17 | 7.87e-19 | 125.82 |

**NOTE:** Effective wavelengths of the filters: UVIT FUV: 1481 Å, UVIT NUV: 2418 Å, DECaLS r-band: 6382.6 Å